\documentclass[a4paper,11pt]{article}
\pdfoutput=1
\usepackage{jheppub}
\usepackage[T1]{fontenc}
\usepackage{empheq}
\usepackage{url}
\usepackage[utf8]{inputenc}
\usepackage[toc,page]{appendix}
\usepackage[autostyle]{csquotes}
\usepackage{slashed}
\usepackage{graphicx}
\usepackage{longtable}
\usepackage{tabularx}
\usepackage{amsmath}
\usepackage{cancel}
\usepackage{mathtools}
\usepackage{amsfonts}
\usepackage{amssymb}
\usepackage{braket}
\usepackage{dirtytalk}
\usepackage{hyperref}
\usepackage{cleveref}

\newcommand{\bse}{\begin{subequations}}
	\newcommand{\ese}{\end{subequations}}
\numberwithin{equation}{section}

\title{Logarithmic corrections to the entropy of non-extremal black holes in $\mathcal{N}=1$ Einstein-Maxwell supergravity}
\author{Gourav Banerjee}
\author{and Binata Panda}
\affiliation{Department of Physics,\\ Indian Institute of Technology (Indian School of Mines),\\ Dhanbad, Jharkhand-826004, INDIA}
\emailAdd{gourav\_9124@ap.ism.ac.in}
\emailAdd{binata@iitism.ac.in}

\abstract{We reviewed the field redefinition approach of Seeley-DeWitt expansion for the determination of Seeley-DeWitt coefficients from \href{https://arxiv.org/abs/1505.01156}{arXiv:1505.01156}. We apply this approach to compute the first three Seeley-DeWitt coefficients for \say{non-minimal} $\mathcal{N}=1$ Einstein-Maxwell supergravity in four dimensions. Finally, we use the third coefficient for the computation of the logarithmic corrections to the Bekenstein-Hawking entropy of non-extremal black holes following \href{https://arxiv.org/abs/1205.0971}{arXiv:1205.0971}. We determine the logarithmic corrections for non-extremal Kerr-Newman, Kerr, Reissner-Nordstr\"{o}m and Schwarzschild black holes in \say{non-minimal} $\mathcal{N}=1$, $d=4$ Einstein-Maxwell supergravity.}

\begin{document}
	\maketitle
	\flushbottom
\section{Introduction}\label{intro}
  Searching for a consistent theory of quantum gravity is always one of the fascinating topics over the decades. Black holes provide us a platform where both gravity and quantum theory become significant. This property plays an important role in the unification of general relativity and quantum mechanics in a single framework.
  In this context, black hole entropy proves to be an efficient tool  for testing any strong candidate for quantum gravity. 
   Bekenstein and Hawking et al. \cite{Bekenstein:1973ur,Hawking:1974sw,Bardeen:1973gs} first proposed that  black holes can
    have a thermodynamical analogy and hence radiate thermal radiation quantum mechanically like a black body. They behave like thermal bodies with definite entropy called the Bekenstein-Hawking entropy. This entropy is universal and proportional to the area of the horizon of the black hole under consideration. However, it does not depend upon the matter content and their couplings with the black hole background. Moreover, this entropy relation is valid for macroscopically large black holes ($A_{H}\ge l_{p}^{2}$, where $l_{p}$ is plank length) and limited up to the classical regime with two derivative terms. But, considering the higher derivative terms in a general theory of gravity and quantum effects, this entropy demands corrections. Several attempts have been made to correct the area law, adding namely quantum corrections \cite{Bhattacharyya:2012ss,Karan:2019gyn,Banerjee:2010qc,Banerjee:2011jp,Sen:2012rr,Sen:2011ba,Keeler:2014bra,Sen:2012dw,Charles:2015nn,Castro:2018hsc,Karan:2020njm,Karan:2021teq} and higher derivative corrections \cite{Iyer:1994ys,Reall:2019sah,Charles:2016wjs,Wald:1993nt}. In this work, we are predominantly interested in quantum corrections to  Bekenstein-Hawking entropy of a black hole.
      The quantum corrected Bekenstein-Hawking entropy of black holes with large charges can be expressed as,

\begin{equation}\label{quantum corrected entropy}
S_{\text{BH}}=\frac{A_{H}}{4G_{N}}+\mathcal{C}\ln \frac{A_{H}}{ G_{N}}+\text{constant}+\mathcal{O}(A_{H}^{-1}),
\end{equation}
where $A_{H}$ is the horizon area of the black hole and $G_{N}$ is Newton's constant. In \cref{quantum corrected entropy}, the first term represents  the classical Bekenstein-Hawking entropy, $\mathcal{C}\ln\frac{A_{H}}{G_{N}}$ is the logarithmic correction term and $\mathcal{O}(A_{H}^{-1})$ give power-law corrections.

In this work, our primary focus is on the determination of the 
logarithmic correction term to Bekenstein-Hawking entropy \eqref{quantum corrected entropy} for a particular class of black holes. This term is proportional to logarithmic of the area of the black hole with proportionality constant $\mathcal{C}$. Logarithmic corrections are leading order quantum corrections, which arise from the loops of massless fields and their coupling to the black hole background \cite{Bhattacharyya:2012ss,Karan:2019gyn,Banerjee:2010qc,Banerjee:2011jp,Sen:2012rr,Sen:2011ba,Keeler:2014bra,Sen:2012dw,Charles:2015nn,Castro:2018hsc,Karan:2020njm,Karan:2021teq}. These corrections are universal and appear in any gravitational theory. They can be  evaluated by computing the quantum determinant over fluctuating massless fields present in the black hole background. Moreover, these corrections depend  only on the knowledge of infrared physics (low energy data) and do not require the details of ultraviolet completion of the theory. 

Logarithmic corrections to the black hole entropy provide non-trivial information about the microstates of the black holes.
Hence computations of these corrections to the entropy of black holes 
in various supergravity theories have become a rich arena to explore in both microscopic and macroscopic sectors. References \cite{Mandal:2010cj,Sen:2014aja} present an excellent comparative study of macroscopic and microscopic entropy corrections of various black holes in different supergravity theories. By successfully comparing the logarithmic corrections to the entropy of a black hole between both sectors (microscopic and macroscopic), one can test a consistent theory of quantum gravity.  Our work mainly deals with the macroscopic (infrared) part for the determination of the logarithmic corrections. 

 Supergravity theories being the low energy limit of superstring theory, provide a general background to study logarithmic corrections to the Bekenstein-Hawking entropy of a black hole.
  The logarithmic corrections to the entropy of various extremal and non-extremal black holes in different supersymmetric and non-supersymmetric theories have been computed using the Euclidean gravity approach \cite{Bhattacharyya:2012ss,Karan:2019gyn,Banerjee:2010qc,Banerjee:2011jp,Sen:2012rr,Sen:2011ba,Keeler:2014bra,Sen:2012dw,Charles:2015nn,Castro:2018hsc,Karan:2020njm,Karan:2021teq}. However, in this progress, cases of extremal black holes are tremendously explored compared to the non-extremal ones. Higher dimensional non-extremal black hole solutions studied in \cite{Mohaupt:2010fk,Cai:2007ik,Cvetic:1995dn,Myers:1986un} are found to be very interesting. In \cite{Cai:2007ik}, the entropy function for non-extremal black hole was studied. Again, the logarithmic corrections to the entropy of various black holes in non-extremal regime were studied in various literature \cite{Solodukhin:1994yz,Solodukhin:1994st,Fursaev:1994te,Solodukhin:2011gn,Sen:2012dw,Charles:2015nn,Castro:2018hsc}. The present paper is greatly motivated by the works of Sen\cite{Sen:2012dw} and Larsen \cite{Charles:2015nn,Castro:2018hsc}. In \cite{Sen:2012dw} and \cite{Charles:2015nn}, the authors have determined the logarithmic corrections to the entropy of non-extremal Kerr-Newman family\footnote{ Kerr-Newman family includes Kerr-Newman, Kerr, Reissner-Nordstr\"{o}m and Schwarzschild black holes.} of black holes in non-supersymmetric and $\mathcal{N}\ge2$ supergravity theories. The results for logarithmic corrections to the entropy of non-extremal black holes in various supergravity theories may provide directions for studying radiations and other thermal properties of black holes in non-extremal regimes.

 
 As already well known, a non-extremal black hole posses a finite temperature called Hawking temperature. These black holes are not stable and radiate thermal radiation like a black body. As a result, they loose energy continuously. Extremal black holes are a limiting case of non-extremal ones, defined with zero temperature and cease to hawking radiate. Non-extremal black holes are neither restricted by any sort of constraint applied over to their inner and outer horizons nor possess any specific geometry having $AdS_2$ factor. So, unlike the extremal case, one can not use quantum entropy function formalism \cite{Sen:2008yk,Sen:2009vz,Sen:2008vm} to determine the logarithmic corrections for the non-extremal black holes.
  In the current work, we will follow the approach of Sen \cite{Sen:2012dw}\footnote{We have reviewed the approach of determination of logarithmic corrections of non-extremal black holes from \cite{Sen:2012dw} and explicitly presented in \cref{section 5}.} for the determination of logarithmic corrections of black holes in non-extremal limit. A similar  approach is used in \cite{Solodukhin:1994yz,Solodukhin:1994st,Fursaev:1994te,Solodukhin:2011gn} for the determination of quantum corrections to  black hole entropy. Later, Larsen et al. also utilized the approach of \cite{Sen:2012dw} in \cite{Charles:2015nn,Castro:2018hsc} for the determination of logarithmic corrections to the non-extremal black holes in different supergravity theories.

  
The aforementioned developments about the determination of logarithmic corrections to non-extremal black holes in various $\mathcal{N}\ge2$ supergravity theories and $\mathcal{N}=0$ non-supersymmetric theory motivate us to give a closer look at black holes in $\mathcal{N}=1$ supergravity. It will be interesting to
compute the logarithmic corrections to the entropy of non-extremal Kerr-Newman family of black holes in this theory. This will provide another example to study the logarithmic correction to Bekenstein-Hawking entropy in this series of supergravity theories in non-extremal limit, and is also important in search of testing a consistent theory of quantum gravity.
The attractor mechanism of black holes in $\mathcal{N}=1$ supergravity was studied by Ferrara et al. \cite{Andrianopoli:2007rm}, where black holes were presented in the context of $\mathcal{N}=1$ supergravity theory. We perform the logarithmic correction analysis utilizing the heat kernel technique in \say{non-minimal} $\mathcal{N}=1$ Einstein-Maxwell supergravity theory (EMSGT) in four dimensions. In \say{non-minimal} $\mathcal{N}=1$, $d=4$ EMSGT, the vector multiplet is non-minimally coupled with pure $\mathcal{N}=1$ supergravity multiplet. This theory is well studied by Ferrara et al. and the corresponding supergauged action is constructed in \cite{Ferrara}.  The Kerr-Newman family of black holes are the solutions of Einstein-Maxwell system, and its configuration can be described by $\mathcal{N}=1$ supergravity theory coupled with a vector multiplet.

In our earlier work \cite{Banerjee:2020wbr}, we have determined the logarithmic corrections of Kerr-Newman family of extremal black holes in \say{non-minimal} $\mathcal{N}=1$, $d=4$ EMSGT using quantum entropy function formalism \cite{Sen:2008yk,Sen:2009vz,Sen:2008vm}. Apart from this, the determination of logarithmic corrections in different classes of $\mathcal{N}=1$, $d=4$ EMSGTs (namely $\mathcal{N}=1$ theory as truncation of $\mathcal{N}=2$ theory and \say{minimally coupled} $\mathcal{N}=1$ theory) has been carried out in \cite{Ferrara:2011qf,Karan:2020njm}. All these above works are mainly focused on determining the entropy corrections of extremal black holes. The work \cite{Karan:2020njm} by Karan et al. also presents logarithmic correction  results for the Kerr-Newman family of black holes in the non-extremal regime, in a particular class of $\mathcal{N}=1$ theory obtained from multiplet decomposition of $\mathcal{N}=2$ theory. This $\mathcal{N}=1$ theory obtained from truncation is an approximated theory, whose multiplet decomposition is done following some set assumptions.\footnote{Kindly refer section 4.1 of \cite{Ferrara:2011qf} for the required assumptions used in this multiplet decomposition.} On the contrary, in the current work we have considered   a more general \say{non-minimal} $\mathcal{N}=1$, $d=4$ EMSGT. It is different from the $\mathcal{N}=1$ theory obtained from \say{truncation} of $\mathcal{N}=2$ theory in the sense that it does not follow any sort of those assumptions considered while multiplet decompositions. We have determined the logarithmic corrections to the entropy of Kerr-Newman, Kerr, Reissner-Nordstr\"{o}m and Schwarzschild black holes in the non-extremal regime  in this \say{non-minimal} $\mathcal{N}=1$, $d=4$ EMSGT. It will be interesting to observe how the entropy of black holes changes from extremal \cite{Banerjee:2020wbr} to non-extremal limit within the same theory. This will also be helpful to develop the microstate picture of the black hole in the non-extremal regime to this theory.

The logarithmic corrections to the entropy of black holes are computed using the heat kernel analysis of one-loop effective action of a theory\cite{Bhattacharyya:2012ss,Karan:2019gyn,Banerjee:2010qc,Banerjee:2011jp,Sen:2012rr,Sen:2011ba,Keeler:2014bra,Sen:2012dw,Charles:2015nn,Castro:2018hsc,Karan:2020njm,Karan:2021teq}. General Seeley-DeWitt expansion approach  \cite{Vassilevich:2003ll} within heat kernel is the most sophisticated approach\footnote{Please see the references of \cite{Banerjee:2020wbr} for other approaches of heat kernel.} for the analysis of one-loop effective action due to its non-dependency in background geometry as well as supersymmetry. This general Seeley-DeWitt expansion approach has been followed in most of the earlier works \cite{Bhattacharyya:2012ss,Karan:2019gyn,Banerjee:2020wbr,Karan:2017txu,Karan:2020njm,Karan:2021teq} for the computation of logarithmic corrections to black hole entropy. Despite being a straightforward approach, the computations of Seeley-DeWitt coefficients from one-loop fluctuated action by the general approach \cite{Vassilevich:2003ll} is very much tedious and tiresome for the theories having a complicated form of quadratic fluctuated action. We experienced these issues particularly while dealing with the computation of Seeley-DeWitt coefficients for minimal $\mathcal{N}=2$, $d=4$ \cite{Karan:2019gyn}, matter coupled $\mathcal{N}\ge1$ \cite{Karan:2020njm}, \say{non-minimal} $\mathcal{N}=1$, $d=4$ EMSGTs \cite{Banerjee:2020wbr} and generalized Einstein-Maxwell theory \cite{Karan:2021teq}. To tackle such extensive and complicated calculations, here we have adopted the field redefinition Seeley-DeWitt expansion approach to determine the corresponding coefficients in this article. This field redefinition approach of Seeley-DeWitt expansion was put forward by Larsen et al. in \cite{Charles:2015nn,Castro:2018hsc}. It is a customized form of general Seeley-DeWitt expansion \cite{Vassilevich:2003ll}. This customization is based on introducing a redefined field into the quadratic fluctuated action in order to simplify the complicated one-loop action. 
 Rest of the steps to determine the Seeley-DeWitt coefficients remain the same as that of the general approach\cite{Vassilevich:2003ll}.
 Although this approach provides a short and simplified computation, it has its own limitations. One has to know exactly about the type of field redefinition to be introduced in a theory as it depends upon the choice of fields present in theory.  Only that field redefinition is considered for which the fluctuated action takes a simplified form. Hence it causes a loss of generality and also makes it impossible to keep track of individual original field contributions throughout the computations.
 
The present work is greatly motivated by the success in the computation of Seeley-DeWitt coefficients for minimal $\mathcal{N}=2$, $d=4$ EMSGT through the field redefinition Seeley-DeWitt expansion  approach \cite{Charles:2015nn}. The results of \cite{Charles:2015nn} perfectly matched with the Seeley-DeWitt coefficients for the same theory obtained through the general approach in \cite{Bhattacharyya:2012ss,Karan:2019gyn,Karan:2020njm}. So, the field redefinition approach of Seeley-DeWitt expansion \cite{Charles:2015nn} provided us a platform to check the consistency of the results for Seeley-DeWitt coefficients for \say{non-minimal} $\mathcal{N}=1$, $d=4$ EMSGT. These coefficients have already been computed in our earlier work \cite{Banerjee:2020wbr}, but following the general approach of Seeley-DeWitt expansion \cite{Vassilevich:2003ll}.  We first reviewed the computation of the first three Seeley-DeWitt coefficients for the bosonic gravity multiplet of $\mathcal{N}=2$, $d=4$ EMSGT from \cite{Charles:2015nn} as it coincides with the bosonic part of \say{non-minimal} $\mathcal{N}=1$, $d=4$ EMSGT. For the fermionic sector, we introduced a new field, which is defined with a particular form of field redefinition \eqref{field redefinition fermion}, and computed the first three Seeley-DeWitt coefficients for this sector as well. The results are found to be in perfect agreement with our earlier result\cite{Banerjee:2020wbr}.

The paper is organized as follows. \Cref{section 2} presents a general review on analysis of  effective action through heat kernel expansion. Here, we first expressed the relationship between the one-loop effective action and Seeley-DeWitt coefficients. Then, we presented the methodology of the field redefinition Seeley-DeWitt expansion approach \cite{Charles:2015nn} to compute the Seeley-DeWitt coefficients. In \cref{section 4}, we start with a discussion on the  general properties of \say{non-minimal} $\mathcal{N}=1$, $d=4$ EMSGT under consideration. We presented the equations of motion and identities for the action of the concerned theory in \ref{section 3}. Then, we applied the field redefinition Seeley-DeWitt expansion approach \cite{Charles:2015nn} to \say{non-minimal} $\mathcal{N}=1$, $d=4$ EMSGT, and computed the first three Seeley-DeWitt coefficients for this theory. The results are in accordance with those in our earlier work \cite{Banerjee:2020wbr}. \Cref{section 5} begins with a review of the approach of \cite{Sen:2012dw} for the determination of logarithmic corrections to the entropy of non-extremal black holes using a particular Seeley-DeWitt coefficient in arbitrary black hole background. We then applied this framework to Kerr-Newman metric and computed the logarithmic corrections to the Bekenstein-Hawking entropy of non-extremal Kerr-Newman, Kerr, Reissner-Nordstr\"{o}m and Schwarzschild black holes in \say{non-minimal} $\mathcal{N}=1$, $d=4$ EMSGT. The results are presented in \cref{entropy KN,entropy K,entropy RN,entropy S}. These results are new as well as unique and will provide a testing ground for any microscopic approach to the same problem in the non-extremal limit. Finally, we conclude in \cref{section 6} with a discussion about all the results obtained. We have presented some details of the calculations in \Cref{Appendix}. 
\section{Seeley-DeWitt expansion of heat kernel via field redefinition approach}\label{section 2}
In this section, we start with a general discussion of Seeley-DeWitt expansion of heat kernel to analyze the one-loop effective action in a 4D field theory. The working procedure evaluates the functional determinant of quadratic fluctuated action, which encapsulates all the information about the one-loop effective action.
Then, we present a brief review of the field redefinition approach of computation of Seeley-DeWitt coefficients from \cite{Charles:2015nn}.

\subsection{Heat kernel analysis of effective action}

Let's consider the partition function $\mathcal{Z}$ in a four-dimensional compact, smooth Riemannian manifold at finite temperature $T$. The manifold is associated with arbitrary fields $\varphi_{m}$. Then, the partition function is given by
\begin{equation}\label{partition function}
\mathcal{Z}=\int [D\varphi_{m}]\text{exp}(-\mathcal{S}[\varphi_{m}]).
\end{equation}
$[D\varphi_m]$ is a functional integral overall configuration of arbitrary fields $\varphi_{m}$ and $\mathcal{S}$ is Euclidean action. The integral \eqref{partition function} is evaluated using saddle point approximation  around the stationary saddle points, which satisfy  the classical equations of motion of a theory.
 In order to evaluate the one-loop correction to the partition function, we begin by following field expansion around the background:
\begin{equation}\label{field fluctuations}
\varphi_{m}=\bar{\varphi}_{m}+\tilde{\varphi}_{m},
\end{equation}
where $\tilde{\varphi}_{m}$ is set of quantum fields fluctuated around the classical background field $\bar{\varphi}_{m}$. These background fields $\bar{\varphi}_m$ correspond to the stationary saddle points. Expanding the action up to quadratic order by imposing the fluctuations \eqref{field fluctuations}, we get the following quadratic fluctuated action:    
\begin{equation}\label{quadratic fluctuated action}
\mathcal{S}_{2}=\int d^4x \sqrt{\bar{g}}\tilde{\varphi}_{m}\Lambda^{mn}\tilde{\varphi}_{n},
\end{equation}
 $\Lambda$ in \cref{quadratic fluctuated action} is the kinetic operator acting over quantum fields.
 The form of $\Lambda$ should be Hermitian, Laplacian,\footnote{ For Fermionic fields in the theory, the operator is required to be put in Laplacian form following \cref{K2}.} minimal second order, pseudo-differential type as per the requirement of heat kernel analysis \cite{Karan:2019gyn}. The one-loop effective action $W$ is related to the kinetic operator $\Lambda$ as \cite{Castro:2018hsc}
\begin{equation}\label{identity effective action}
\exp(-W)=\int [D\tilde{\varphi}_{m}]\exp(-\int d^4 x \sqrt{\bar{g}} \tilde{\varphi}\Lambda\tilde{\varphi})={\det}^{-\chi/2}\Lambda.
\end{equation}
$\chi$ is +1 for bosons and -1 for fermions. To study the spectrum of $\Lambda$, we introduce heat kernel $K(x,y;s)$, which is the solution of the standard heat equation. Here, $x$ and $y$ are points on the manifold, whereas $s$ is proper time coordinate, treated as the heat kernel parameter. Setting $x=y$, we define a new quantity heat trace $D(s)$,
\begin{equation}\label{ds}
D(s)=\int d^4x \sqrt{\bar{g}} K(x,x;s).
\end{equation}
 Then, the one-loop effective action $(W)$ can be expressed in terms of heat trace $D(s)$ as
\begin{equation}\label{one loop effective action}
W=-\frac{1}{2}\int_{\epsilon}^{\infty}\frac{ds}{s} \chi D(s),
\end{equation}
where
\begin{align}\label{ds1}
	\begin{split}
	D(s)&=\text{tr}\thickspace (e^{-s\Lambda})
	=\sum_{i}e^{-s\lambda_{i}}
	=\sideset{}{'}\sum_{\substack{i\\\lambda_i\ne0}}e^{-s\lambda_{i}}+\int d^4 x \sqrt{\bar{g}} K^{\text{zm}}(x,x;0),
	\end{split}
\end{align}
and $\lambda_{i}$ are the eigenvalues of the kinetic operator $\Lambda$. $\sideset{}{'}\sum$ defines the summation over non-zero values of $\lambda_i$.  $K^{\text{zm}}(x,x;0)$ is zero mode contribution, associated with the modes having zero eigenvalues of kinetic operator in $K(x,x;s)$.  $\epsilon$ is the UV limit in \cref{one loop effective action}.\footnote{We have considered $\epsilon \sim l_{p}^{2} \sim G_{N} \sim \frac{1}{16 \pi}$ throughout the work.}
The heat trace $D(s)$ after small perturbative expansion in proper time is given as 
\begin{equation}\label{expansion}
D(s)=\int d^4 x \sqrt{\bar{g}} \sum_{n=0}^{\infty}s^{n-2}a_{2n}(x).
\end{equation}
This expansion is called the Seeley-DeWitt expansion \cite{DeWitt:1965ff,Vassilevich:2003ll,Duff:1977vv,Christensen:1979ww,Christensen:1980xx,Duff:1980yy,Birrel:1982zz,Gilkey:1984xy,Gilkey:1975cd,Avramidi:1994th,DeWitt:1967gg,DeWitt:1967hh,DeWitt:1967ii,Seeley:1966tt,Seeley:1969uu}. Here $a_{2n}$ is called Seeley-DeWitt coefficients. So, the problem of finding the quantum correction to the theory is related to determining heat trace $D(s)$, and this has to be done by evaluating the Seeley-DeWitt coefficients $a_{2n}$ of the theory. In the next subsection, we will put light on the computational method of these Seeley-DeWitt coefficients $a_{2n}$.
\subsection{Computation of Seeley-DeWitt coefficients by field redefinition Seeley-DeWitt expansion approach}\label{computation of coefficients}
The most common and general approach for the computation of Seeley-DeWitt coefficients was presented by Vassilevich \cite{Vassilevich:2003ll}. Later, Larsen et al. \cite{Charles:2015nn,Castro:2018hsc} customized the general approach \cite{Vassilevich:2003ll} by introducing a field redefinition into the quadratic fluctuated action of a theory. By doing so, the further computations of Seeley-DeWitt coefficients for the theory having complex form of action become short and simplified. Below we present a brief review on computation of the Seeley-DeWitt coefficients through the field redefinition approach of Seeley-DeWitt expansion, following the treatment of \cite{Charles:2015nn}.

Let's consider the quadratic fluctuated action \eqref{quadratic fluctuated action} for any theory for which the form of the kinetic operator $(\Lambda)$ turns out to be complicated, making it difficult to follow the general approach for the computation of Seeley-DeWitt coefficients. Then, the action \eqref{quadratic fluctuated action} is calibrated by introducing a suitable new field $\tilde{\varPhi}$, formulated from the set of original fields $\tilde{\varphi}$ present in the theory (more clearly, please see  \eqref{field redefinition boson} and \eqref{field redefinition fermion}). 
The structure of this redefined field can be guessed from the form of one-loop action of the theory. This redefinition has to be done in such a way that it brings a simplified  form of the kinetic operator $(\Lambda)$ from which the further computation of Seeley-DeWitt coefficients are easily accomplished. 
 However, in the above process, the basic structure of the new redefined form of $\Lambda$ still retains the same form as general $\Lambda$\footnote{Kindly refer \cite{Banerjee:2020wbr,Karan:2019gyn} for an explicit review of the general approach of computation of Seeley-DeWitt coefficients\cite{Vassilevich:2003ll}.} constructed without any field redefinition and is given as

\begin{equation}\label{lambda 1}
\Lambda^{mn}=\pm\lbrace(D^{\rho}D_{\rho})G^{mn}+(N^{\rho}D_{\rho})^{mn}+P^{mn}\rbrace,
\end{equation}   
where $+$ve and $-$ve sign is to be considered for bosonic and fermionic quadratic fluctuated action, respectively. $G$ is the effective metric in field space, which is $\bar{g}^{\mu\nu}$ for vector field, $\bar{g}^{\mu\nu}\mathbb{I}_4$ for Rarita Schwinger field  and $\mathbb{I}_4$ for Dirac field with $\mathbb{I}_4$ being the 4D identity matrix. $N$ and $P$ are arbitrary matrices associated with the derivative and non-derivative part of $\Lambda$ in \cref{lambda 1}. $D_{\rho}$ is an ordinary covariant derivative. Rest of the steps are exactly same as the general approach prescribed in \cite{Vassilevich:2003ll}. \Cref{lambda 1} is to be expressed in the form
\begin{equation}\label{lambda 2}
\Lambda^{mn}=\pm\lbrace(\mathcal{D}^{\rho}\mathcal{D}_{\rho}) \mathbb{I}^{mn}+E^{mn}\rbrace,
\end{equation} 
where
\begin{align}\label{heat kernel parameters}
&\mathcal{D}_{\rho}=D_{\rho}+\omega_{\rho}, \enspace \mathbb{I}=G, \enspace \omega_{\rho}=\frac{1}{2}N_{\rho},\enspace E=P-\omega^{\rho}\omega_{\rho}-D^{\rho}\omega_{\rho}.
\end{align} 
$\mathcal{D_{\rho}}$ is the new effective covariant derivative, $\omega_{\rho}$ is the field connection. The field strength $\Omega_{\alpha \beta}$ associated with curvature $\mathcal{D}_{\rho}$ as
\begin{equation}\label{Omega field strength}
\Omega_{\alpha \beta}\equiv[\mathcal{D}_{\alpha},\mathcal{D}_{\beta}].
\end{equation}
 The Seeley-DeWitt coefficients in terms of $\mathbb{I}$, $E$ , $\Omega_{\alpha \beta}$ and other local background invariant parameters are given by relation \cite{Vassilevich:2003ll,Gilkey:1984xy,Gilkey:1975cd}:
\begin{align}\label{a 2n}
\begin{split}
\chi (4\pi)^2 a_0(x)= &\text{tr}~(\mathbb{I}),
\\
\chi (4\pi)^2 a_2(x)=&\frac{1}{6}\text{tr}~(6E+R \mathbb{I}),
\\
\chi (4\pi)^2 a_4(x)=&\frac{1}{360}\text{tr}~\big\lbrace 60 R E +180 E^2+30 \Omega^{\mu\nu}\Omega_{\mu\nu}
\\
&\quad+(5R^2-2R^{\mu\nu}R_{\mu\nu}+2R^{\mu\nu\rho\sigma}R_{\mu\nu\rho\sigma})\mathbb{I}\big\rbrace,
\end{split}
\end{align}
where $R$, $R_{\mu\nu}$ and $R_{\mu\nu\rho\sigma}$ are the usual curvature tensor associated with background metric. In expressions of Seeley-DeWitt coefficients \eqref{a 2n}, we have ignored the total derivative terms because we are particularly interested in the manifold having no boundary.

 We summarize the field redefinition Seeley-DeWitt expansion  approach \cite{Charles:2015nn} mentioned above for the computation of Seeley-DeWitt coefficients of a theory in the following algorithm.
\begin{enumerate}
\item Fluctuate the fields around the classical background \eqref{field fluctuations} and obtain the quadratic fluctuated action \eqref{quadratic fluctuated action}.
\item Gauge fix the quadratic fluctuated action \eqref{quadratic fluctuated action} by adding the proper gauge fixing term in action obtained in step 1. Thus, we have a gauge fixed quadratic fluctuated action.
\item Express the gauge fixed quadratic fluctuated action obtained in step 2 by inserting a proper redefined field, so that the new redefined quadratic fluctuated action takes a simplified form than obtained in step 2.
\item  Identify the redefined kinetic differential operator $\Lambda_{}$ (which is associated with simplified action obtained in step 3) in the form \eqref{lambda 1}. Also, find associated ghost fields arose due to gauge fixing in step 2. The ghost sector has to be considered separately. 
\item Identify the form of $G$, $N_{\rho}$ and $P$ from the calibrated $\Lambda$ having no ghost.
\item This form of $\Lambda_{}$ obtained in step 4 is made to fit in the prescription of \cref{lambda 2}. From where identify the expression of $\mathbb{I}$, $E$ and $\omega_{\rho}$ using \cref{heat kernel parameters}. The expression of $\Omega_{\alpha \beta}$ can be obtained from \cref{Omega field strength}.
\item Now, compute the traces of $\mathbb{I}$, $E$, $E^2$ and $\Omega_{\alpha \beta}\Omega^{\alpha \beta}$, also make use of equations of motion and other identities in this trace calculations for necessary simplifications.
\item Use the trace results obtained in step 7  to compute the first three Seeley-DeWitt coefficients utilizing \cref{a 2n}.
\item Now similarly compute the Seeley-DeWitt coefficients associated with the ghost sector following above-mentioned steps 4-8. Most often, for the ghost action, the form of $\Lambda$ turns out to be in simplified form. Hence one need not introduce any redefined field for ghosts and can follow the general approach \cite{Vassilevich:2003ll} to compute the  corresponding coefficients.

\item Finally, add both the Seeley-DeWitt coefficients, obtained in step 8 and step 9  to find the total Seeley-DeWitt coefficients for the theory.

\end{enumerate}
Kindly note these Seeley-DeWitt coefficients are defined with local background invariants, i.e., the curvatures tensors associated with the background metric. In the next section, we would discuss about the \say{non-minimal} $\mathcal{N}=1$ EMSGT in four dimensions. Subsequently, we determine Seeley-DeWitt coefficients $a_0$, $a_2$ and $a_4$ for this theory following the above steps (1-10).

\section{Seeley-DeWitt coefficients in ``non-minimal" $\mathcal{N}=1$, $d=4$ EMSGT }\label{section 4}

We apply the  methodology outlined in \cref{section 2} for computation of Seeley-DeWitt coefficients to \say{non-minimal} $\mathcal{N}=1$, $d=4$ EMSGT and compute  the first three Seeley-DeWitt coefficients.
As already discussed in \cref{section 2}, we prefer to carry out this exercise via the field redefinition approach of Seeley-DeWitt expansion \cite{Charles:2015nn}. We shall consider  the bosonic and fermionic fields of this theory separately. We review the bosonic sector, described by  \cref{non fluctuated bosonic action}, following 
\cite{Charles:2015nn}.  We then compute the coefficients for the fermionic sector by introducing a proper redefined field. Finally, we determine the total Seeley-DeWitt coefficients for the theory by adding the contributions from both sectors \textemdash bosonic and fermionic.  In the later section, we use the third Seeley-DeWitt coefficient to determine the logarithmic corrections to the entropy of non-extremal black holes in this theory.

\subsection{``Non-minimal" $\mathcal{N}=1$, $d=4$ EMSGT}\label{section 3}

The \say{non-minimal} $\mathcal{N}=1$ EMSGT in four dimensions is characterized with field contents: spin 2 graviton $(g_{\mu\nu})$, spin 1 gauge field $(A_{\mu})$ with corresponding superpartners spin 3/2 gravitino $(\psi_{\mu})$ and spin 1/2 gaugino field $(\lambda)$.\footnote{ For more details on  \say{non-minimal} $\mathcal{N}=1$, $d=4$ EMSGT, please refer \cite{Banerjee:2020wbr}.} The action corresponding to this theory is constructed by Ferrara et al. \cite{Ferrara} by coupling pure $\mathcal{N}=1$ supergravity multiplet $(g_{\mu\nu},\psi_{\mu})$ with vector multiplet $(A_{\mu}, \lambda)$. Here the gaugino field of vector multiplet interacts non-minimally with the gravitino field of supergravity multiplet, and only the gauge field is minimally coupled to gravity in the supergauged action of this theory. Our motive is to determine the logarithmic corrections to the entropy of black holes by heat kernel analysis in this \say{non-minimal} $\mathcal{N}=1$, $d=4$ EMSGT. Below we present the classical equations of motion and concerned identities for this theory.

We begin with the action for \say{non-minimal} $\mathcal{N}=1$, $d=4$ EMSGT \cite{Ferrara},
\begin{equation}\label{action}
\mathcal{S}=\int d^4 x \sqrt{g}\mathcal{L}_{\text{EM}}+\int d^4x \sqrt{g} \mathcal{L}_{f},
\end{equation}
where
\begin{equation}\label{non fluctuated bosonic action}
\mathcal{L}_{\text{EM}}=\mathcal{R}-F^{\mu\nu}F_{\mu\nu},
\end{equation}
and
\begin{equation}\label{fermionic action}
\mathcal{L}_{f}=-\frac{1}{2} \bar{\psi}_{\mu}\gamma^{[\mu\rho\nu]}D_{\rho}\psi_{\nu}-\frac{1}{2}\bar{\lambda}\gamma^{\rho}D_{\rho}\lambda+\frac{1}{2\sqrt{2}}\bar{\psi}_{\mu} F_{\alpha \beta}\gamma^{\alpha}\gamma^{\beta}\gamma^{\mu}\lambda. 
\end{equation}
 $\mathcal{R}$, $F_{\mu\nu}=\partial_{[\mu}A_{\nu]}$ and $\gamma$ are  Ricci scalar curvature, gauge field strength and Majorana gamma spinor matrices, respectively in \cref{action}. $\gamma^{[\mu\rho\nu]}$ is antisymmetric product of gamma matrices. $\mathcal{L}_{\text{EM}}$ and $\mathcal{L}_{f}$ denote bosonic and fermionic lagrangian of \say{non-minimal} $\mathcal{N}=1$, $d=4$ EMSGT, respectively. To have a locally supersymmetric action, one needs to include the four-fermion interaction terms in the action part of \eqref{fermionic action} \cite{Ferrara}. However, the heat kernel analysis is to be done on the quadratic fluctuated action for the determination of logarithmic corrections to the entropy of black holes. The one-loop contributions from four-fermion interaction terms vanish because of the vanishing of background fermions in the Einstein-Maxwell system. So, neglecting the four-fermion interaction terms in \eqref{fermionic action}  will not affect the form of quadratic fluctuated action of \say{non-minimal} $\mathcal{N}=1$, $d=4$ EMSGT. 

An arbitrary solution $(\bar{g}_{\mu\nu},\bar{A}_{\mu})$ satisfying the classical equations of motion of Einstein-Maxwell theory is embedded in four dimensional $\mathcal{N}=1$ supersymmetric theory. This geometry satisfies
\begin{equation}\label{EOM}
R_{\mu\nu}=2\bar{F}_{\mu\rho}\bar{F_{\nu}}^{\rho}-\frac{1}{2}\bar{g}_{\mu\nu}\bar{F}^{\rho\sigma}\bar{F}_{\rho \sigma},
\end{equation}
\begin{equation}
D^{\mu}\bar{F}_{\mu\nu}=0, \qquad D_{[\mu}\bar{F}_{\nu\rho]}=0, \qquad R_{\mu[\nu\theta\phi]}=0.
\end{equation}
Field strength $\bar{F}_{\mu\nu}$ in \cref{EOM} is associated with background gauge field $\bar{A}_{\mu}$. The trace of Einstein equation for the theory gives $R=0$, representing a classical solution. These constraints become very helpful in expressing the coefficients in terms of local invariants. 
\subsection{Seeley-DeWitt coefficients for bosonic sector}\label{bosons}
Here we are particularly focused on the computation of the first three Seeley-DeWitt coefficients for the bosonic sector of \say{non-minimal} $\mathcal{N}=1$, $d=4$ EMSGT. The bosonic sector of this theory includes only two kinds of fields: $g_{\mu\nu}$ and $A_{\mu}$. The corresponding Seeley-DeWitt coefficients were computed in various works by the general approach \cite{Bhattacharyya:2012ss,Karan:2019gyn,Karan:2021teq} and by field redefinition approach \cite{Charles:2015nn}. In this subsection, we briefly review this computation of the coefficients following field redefinition approach. Kindly note that we consider a different scaling of field strength tensor $F_{\mu\nu}$ than that of \cite{Charles:2015nn}. The final results of this section, although already known, provide a consistency check on the results obtained in \cite{Bhattacharyya:2012ss,Karan:2019gyn} and help us to set our conventions for the rest of the paper.
  
We begin by  constructing the quadratic fluctuated action from \eqref{non fluctuated bosonic action} by imposing the following fluctuations on the fields present in theory:
\begin{equation}
g_{\mu\nu}=\bar{g}_{\mu\nu}+\sqrt{2}h_{\mu\nu},\enspace A_{\mu}=\bar{A}_{\mu}+\frac{1}{2}a_{\mu},
\end{equation}
where $h_{\mu\nu}$ and $a_{\mu}$ are the corresponding fluctuations in the background fields $\bar{g}_{\mu\nu}$ and $\bar{A}_{\mu}$ respectively.
We gauge-fix the quadratic fluctuated action by adding the following gauge fixing term\cite{Bhattacharyya:2012ss}
\begin{equation}
-\int d^4 x \sqrt{\bar{g}}\Big\lbrace\Big(D^{\mu}h_{\mu\rho}-\frac{1}{2}D_{\rho} h\Big)\Big(D^{\nu}{h_{\nu}}^\rho-\frac{1}{2}D^{\rho}h\Big)+\frac{1}{2}(D^{\mu}a_{\mu})(D^{\nu}a_{\nu})\Big\rbrace,
\end{equation}
where we have denoted ${h=h^{\mu}}_{\mu}=\text{tr}\thickspace (h_{\mu\nu})$.  We thus obtain the gauge-fixed quadratic fluctuated action excluding the ghost part\cite{Bhattacharyya:2012ss},
\begin{align}\label{fluctuatd EM action}
\begin{split}
\mathcal{S}_{2}&=\frac{1}{2}\int d^4 x \sqrt{\bar{g}}\Big\lbrace-h^{\mu\nu} \Delta h_{\mu\nu}+a_{\mu} (D^{\rho}D_{\rho})^{\mu\nu}a_{\nu}-a_{\mu}R^{\mu\nu}a_{\nu}
\\
& -\frac{1}{2}\bar{F}_{\mu\nu}\bar{F}^{\mu\nu}
\big(h^2-2h^{\rho \sigma}h_{\rho \sigma}\big)-4\bar{F}_{\mu\nu} \bar{F}_{\rho \sigma} h^{\mu\rho}h^{\nu\sigma}-8 \bar{F}_{\mu \rho} \bar{F_{\nu}}^{\rho} {h^{\mu}}_{\sigma} h^{\nu \sigma}
\\
&+4\bar{F}_{\mu \rho} \bar{F_{\nu}}^{\rho} h h^{\mu\nu}-\sqrt{2}\bar{F}^{\mu\nu} h f_{\mu\nu}+4\sqrt{2} \bar{F}^{\mu\nu} {h_{\nu}}^{\rho}f_{\mu\rho} \Big\rbrace,
\end{split}
\end{align}
where
\begin{equation}
\begin{split}
\Delta h_{\mu\nu}=&-\Box h_{\mu\nu}-R_{\mu \tau} {h^{\tau}}_{\nu}-R_{\nu\tau}{h_{\mu}}^{\tau}-2R_{\mu\rho\nu\tau} h^{\rho \tau}+\frac{1}{2}\bar{g}_{\mu\nu}\bar{g}^{\rho \sigma}\Box h_{\rho \sigma}
\\
&+R \thickspace h_{\mu\nu}+(\bar{g}_{\mu\nu}R^{\rho \sigma}+R_{\mu\nu} \bar{g}^{\rho\sigma})h_{\rho \sigma}-\frac{1}{2} R \bar{g}_{\mu\nu} \bar{g}^{\rho \sigma}h_{\rho \sigma}\label{h},
\end{split}
\end{equation}
 $f_{\mu\nu}\equiv D_{[\mu}a_{\nu]}$, $\Box\equiv D^{\rho}D_{\rho}$ in \cref{fluctuatd EM action,h}. Now setting $R=0$ and using equations of motion \eqref{EOM}, the bosonic quadratic fluctuated action \eqref{fluctuatd EM action} takes the form
\begin{align}\label{action 2} 
\begin{split}
\mathcal{S}_{2}=&\frac{1}{2}\int d^4 x \sqrt{\bar{g}}\Big\{h^{\mu\nu}\Box h_{\mu\nu}-\frac{1}{2}h\Box h+a^{\alpha} \Box a_{\alpha}-a_{\alpha}R^{\alpha\beta}a_{\beta}
\\
&+2h_{\mu\nu}R^{\mu\alpha\nu\beta}h_{\alpha \beta}-2h_{\mu\nu}R^{\rho \nu}{h^{\mu}}_{\rho}-h^{\alpha\beta} \bar{F}^{\rho \sigma} \bar{F}_{\rho \sigma} h_{\alpha\beta}+\frac{1}{2}\bar{F}^{\rho \sigma}\bar{F}_{\rho \sigma}h^{2}
\\
&-4h_{\mu\nu}\bar{F}^{\mu \alpha} \bar{F}^{\nu \beta} h_{\alpha \beta}-\sqrt{2} h \bar{F}^{\mu\nu} f_{\mu\nu}+4\sqrt{2}{h^{\mu}}_{\nu}\bar{F}^{\nu \rho}f_{\mu\rho} \Big\}.
\end{split}
\end{align}
We need to calibrate the above action to a simplified form as depicted in \cref{computation of coefficients}. We introduce a particular redefined field of the form  \eqref{field redefinition boson}, so that the action \eqref{action 2} and corresponding Laplacian differential operator $\Lambda$ reduce to a comparatively simpler form. The redefined field $\Phi_{\mu\nu}$, obtained by a linear combination of the fields $h_{\mu\nu}$ and $h$, is,

\begin{equation}\label{field redefinition boson}
\Phi_{\mu\nu}=h_{\mu\nu}-\frac{1}{4}\bar{g}_{\mu\nu}h,\>\>\>\>\>\text{where} \>\>\>\text{tr} \thickspace (\Phi_{\mu\nu})=0, 
\end{equation}
 Using this field redefinition \eqref{field redefinition boson}, the bosonic quadratic fluctuated action takes the form \cite{Charles:2015nn}
\begin{equation}\label{action 3}
\mathcal{S}_{2}=\frac{1}{2}\int d^4x \sqrt{\bar{g}} \tilde{\varphi}_{m}\Lambda^{mn}\tilde{\varphi}_{n}, 
\end{equation}
where the redefined form of $\Lambda^{mn}_{}$ is expressed as 
\begin{align}\label{Lambda action 3}
\begin{split}
\tilde{\varphi}_{m}\Lambda^{mn}_{}\tilde{\varphi}_{n}=&\Phi^{\mu\nu}\Box \Phi_{\mu\nu}-\frac{1}{4}h \Box h +a^{\alpha}\Box a_{\alpha}-a_{\alpha}R^{\alpha \beta} a_{\beta}
\\
& +2\Phi_{\mu\nu}R^{\mu\alpha\nu\beta}\Phi_{\alpha \beta}-2\Phi_{\mu\nu}R^{\rho \nu} {\Phi^{\mu}}_{\rho}-\Phi^{\mu\nu}\bar{F}^{\alpha \beta}\bar{F}_{\alpha \beta} \Phi_{\mu\nu}
\\
&-4\Phi_{\mu\nu} \bar{F}^{\mu \alpha} \bar{F}^{\nu\beta}\Phi_{\alpha \beta}-h R^{\mu\nu} \Phi_{\mu\nu} +4\sqrt{2}{\Phi^{\mu}}_{\nu}\bar{F}^{\nu \rho} f_{\mu \rho}.
\end{split}
\end{align}
In order to have a uniform normalization factor of the kinetic term for all the fields in  \cref{Lambda action 3} , let's define a new field $\Theta$ as
\begin{equation}\label{rescale h}
\Theta=-\frac{i}{2}h.
\end{equation}
We also introduce the following effective metric acting on $\Phi_{\mu\nu}$ fields:\footnote{We have defined $G^{mn}$ in \cref{lambda 1}, which is associated with various fields of the bosonic sector as $\tilde{\varphi}_{m}G^{mn}\tilde{\varphi_{n}}=\Phi_{\mu\nu}G^{\mu\nu \thickspace \alpha\beta} \Phi_{\alpha \beta}+a_{\alpha}G^{\alpha \beta}a_{\beta}$ where $G^{\mu\nu \thickspace \alpha\beta}$ is defined in \cref{G 1}, and $G^{\alpha \beta}=\bar{g}^{\alpha \beta}$.}
\begin{equation}\label{G 1}
G^{\mu\nu \thickspace \alpha \beta}=\frac{1}{2}\Big(\bar{g}^{\mu\alpha}\bar{g}^{\nu \beta}+\bar{g}^{\mu \beta}\bar{g}^{\nu \alpha}-\frac{1}{2}\bar{g}^{\mu\nu}\bar{g}^{\alpha \beta}\Big).
\end{equation}
This effective metric in \cref{G 1} will help to contract the symmetric pair of indices for the field $\Phi_{\mu\nu}$.  
Using \cref{rescale h,G 1,Lambda action 3}, one finds  the new form of $\Lambda$ as,
\begin{align}\label{Lambda 4}
\begin{split}
\tilde{\varphi}_{m}\Lambda^{mn}_{}\tilde{\varphi}_{n}
=& G^{\mu\nu \thickspace \alpha \beta} \Phi_{\mu\nu} \Box  \Phi_{\alpha \beta}+\Theta \Box \Theta+ \bar{g}^{\alpha \beta} a_{\alpha}  \Box  a_{\beta}-a_{\alpha} R^{\alpha \beta}a_{\beta}
\\
&-2\Phi_{\mu\nu} R^{\sigma \nu}{\Phi^{\mu}}_{\sigma}-2i \Theta R^{\alpha \beta}\Phi_{\alpha \beta}+2\Phi_{\mu\nu}R^{\mu\alpha\nu\beta}\Phi_{\alpha \beta}
\\
&-\Phi^{\mu\nu}\bar{F}^{\alpha \beta} \bar{F}_{\alpha \beta}\Phi_{\mu\nu}-4\Phi_{\mu\nu} \bar{F}^{\mu\alpha}\bar{F}^{\nu\beta}\Phi_{\alpha \beta}+4\sqrt{2}{\Phi^{\mu}}_{\nu}\bar{F}^{\nu \rho} f_{\mu \rho},
\\
=&\Phi_{\mu\nu}\Big\{G^{\mu\nu\thickspace\alpha\beta}\Box +2R^{\mu\alpha\nu\beta}-2R^{\beta\nu}\bar{g}^{\alpha \mu}-\bar{g}^{\alpha\mu}\bar{g}^{\beta \nu} \bar{F}^{\rho \sigma}\bar{F}_{\rho \sigma}-4\bar{F}^{\mu\alpha}\bar{F}^{\nu \beta}\Big\}\Phi_{\alpha \beta}
 \\
&+\Theta \Box \Theta+ a_{\alpha}\Big\{ \bar{g}^{\alpha \beta} \Box -R^{\alpha \beta}\Big\}a_{\beta}-i\Phi_{\mu\nu}R^{\mu\nu}\Theta-i\Theta R^{\alpha \beta}\Phi_{\alpha \beta}
\\
&+\sqrt{2}\Phi_{\mu\nu}(D^{\mu} \bar{F}^{\alpha \nu})a_{\alpha}+\sqrt{2}a_{\mu} (D^{\alpha}\bar{F}^{\mu \beta})\Phi_{\alpha \beta}
\\
& +2\sqrt{2}\Phi_{\mu\nu}\Big\{-\bar{g}^{\rho \mu} \bar{F}^{\alpha \nu} + \bar{g}^{\alpha \mu} \bar{F}^{\rho \nu}  \Big\}D_{\rho}a_{\alpha}
\\
&+2\sqrt{2}a_{\mu}\Big\{\bar{g}^{\rho \alpha}\bar{F}^{\mu \beta} -\bar{g}^{\mu \alpha} \bar{F}^{\rho \beta} \Big\}D_{\rho}\Phi_{\alpha \beta}.
\end{split}
\end{align}
Kindly note that the form of $\Lambda$ \eqref{Lambda 4} fits into the necessary Laplacian form \eqref{lambda 1}, which is essential for the heat kernel analysis. One can extract the expression of $P$ and $N_{\rho}$ using \cref{Lambda 4,lambda 1}, as follows
\begin{align}\label{P Boson}
\begin{split}
\tilde{\varphi}_{m} P^{mn} \tilde{\varphi}_{n}=&\Phi_{\mu\nu}\Big\{2R^{\mu\alpha\nu\beta}-2R^{\nu \beta} \bar{g}^{\mu \alpha }-\bar{g}^{\mu \alpha } \bar{g}^{\nu \beta} \bar{F}^{\rho \sigma}\bar{F}_{\rho \sigma}-4\bar{F}^{\mu \alpha} \bar{F}^{\nu \beta}\Big\}\Phi_{\alpha \beta}
\\
&-a_{\alpha}R^{\alpha \beta}a_{\beta}-i\Phi_{\mu\nu} R^{\mu\nu} \Theta-i\Theta R^{\alpha \beta} \Phi_{\alpha \beta}
\\
&+\sqrt{2}\Phi_{\mu\nu}D^{\mu}\bar{F}^{\alpha \nu} a_{\alpha}+\sqrt{2}a_{\mu}D^{\alpha}\bar{F}^{\mu \beta} \Phi_{\alpha \beta},
\end{split}
\end{align}   
\begin{equation}
	\begin{split}\label{N boson}
		\tilde{\varphi}_{m}(N_{\rho})^{mn}\tilde{\varphi}_{n}
		=2\sqrt{2}\Phi_{\mu\nu}(\bar{g}^{\mu \alpha }\bar{F_{\rho}}^{\nu} -\bar{g_{\rho}}^{\mu}\bar{F}^{\alpha \nu})a_{\alpha}+2\sqrt{2}a_{\mu}(\bar{g_{\rho}}^{\alpha}\bar{F}^{\mu \beta} -\bar{g}^{\mu \alpha}\bar{{F}_{\rho}}^{\beta} )\Phi_{\alpha \beta}
	\end{split}
\end{equation}
 The expression of $\omega_{\rho}$ using \cref{heat kernel parameters} is given by,
\begin{align}\label{omega boson}
\begin{split}
\tilde{\varphi}_{m} (\omega_{\rho})^{mn} \tilde{\varphi}_{n}=&\frac{1}{2}\tilde{\varphi}_{m}(N_{\rho})^{mn}\tilde{\varphi}_{n}
\\
=&\sqrt{2}\Phi_{\mu\nu}(\bar{g}^{\mu \alpha }\bar{F_{\rho}}^{\nu} -\bar{g_{\rho}}^{\mu}\bar{F}^{\alpha \nu})a_{\alpha}+\sqrt{2}a_{\mu}(\bar{g_{\rho}}^{\alpha}\bar{F}^{\mu \beta} -\bar{g}^{\mu \alpha}\bar{{F}_{\rho}}^{\beta} )\Phi_{\alpha \beta}.
\end{split}
\end{align}
Once $P$ and $\omega_{\rho}$ are determined, the expressions of $\mathbb{I}$ and $E$ are obtained using \cref{P Boson,omega boson,heat kernel parameters}. The required expressions are,
\begin{align}
\begin{split}
\tilde{\varphi}_{m}\mathbb{I}^{mn}\tilde{\varphi}_{n}=&\Phi_{\mu\nu}G^{\mu\nu \thickspace \alpha\beta}\Phi_{\alpha \beta}+\Theta \Theta+a_{\alpha}\bar{g}^{\alpha \beta}a_{\beta},
\\\label{bosonic E}
\tilde{\varphi}_{m}E^{mn}\tilde{\varphi}_{n}=& 2\Phi_{\mu\nu}R^{\mu\alpha\nu\beta}\Phi_{\alpha \beta}+\frac{3}{2}a_{\alpha}\bar{g}^{\alpha \beta} \bar{F}^{\rho \sigma} \bar{F}_{\rho \sigma}a_{\beta}-i\Phi_{\mu\nu}R^{\mu\nu}\Theta
\\
&-i\Theta R^{\alpha \beta}\Phi_{\alpha \beta}+\sqrt{2}\Phi_{\mu\nu} D^{\mu} \bar{F}^{\alpha \nu} a_{\alpha}
+\sqrt{2} a_{\mu} D^{\alpha} \bar{F}^{\mu\beta}\Phi_{\alpha \beta}
\end{split} 
\end{align}
and simultaneously, the field strength $\Omega_{\rho \sigma}$ can also be determined using \cref{Omega field strength}, which is given as
\begin{align}\label{Bosonic Omega}
\begin{split}
\tilde{\varphi}_{m}(\Omega_{\rho \sigma})^{mn} \tilde{\varphi}_{n}=&\tilde{\varphi}_{m}[D_{\rho}, D_{\sigma}]^{mn}\tilde{\varphi}_{n}+\tilde{\varphi}_{m}{D_{[\rho}\omega_{\sigma]}}^{mn}\tilde{\varphi}
_{n}+\tilde{\varphi}_{m}[\omega_{\rho},\omega_{\sigma}]^{mn}\tilde{\varphi}_{n}
\\
 =& \Phi_{\mu\nu}\Big\{2\bar{g}^{\beta \nu}{R^{\mu\alpha}}_{\rho\sigma}+\Big(2 \bar{g_{\sigma}}^{\alpha} \bar{F_{\rho}}^{\nu} \bar{F}^{\mu \beta} -2 \bar{g}^{\mu \alpha} \bar{F_{\rho}}^{\nu} \bar{F_{\sigma}}^{\beta} -2 \bar{g_{\rho}}^{\mu} \bar{F}^{\nu \alpha} \bar{F_{\sigma}}^{\beta} 
\\
&- \bar{g_{\rho}}^{\mu} \bar{g_{\sigma}}^{\alpha} R^{\nu \beta} -\frac{1}{2}\bar{g}^{\nu \beta} \bar{{g}_{\rho}}^{\mu}\bar{{g}_{\sigma}}^{\alpha} \bar{F}^{\theta \phi}\bar{F}_{\theta \phi}-(\rho \leftrightarrow\sigma)
 \Big) \Big\}\Phi_{\alpha \beta}
 \\
 & +a_{\alpha}\Big\{{R^{\alpha \beta}}_{\rho \sigma}+\Big(\bar{{g}_{\rho}}^{\beta} \bar{F}^{\alpha \phi}\bar{F}_{\sigma \phi}+2\bar{F}_{\rho \sigma}\bar{F}^{\beta \alpha}-2 \bar{{F}_{\rho}}^{\beta}\bar{{F}_{\sigma}}^{\alpha}+2\bar{{F}_{\rho}}^{\alpha} \bar{{F}_{\sigma}}^{\beta}
 \\
 &+\bar{g^{\alpha}}_{\sigma} \bar{{F}_{\rho}}^{\phi}\bar{{F}^{\beta}}_{\phi}-(\rho\leftrightarrow\sigma) \Big) \Big\}a_{\beta}-\sqrt{2}\Phi_{\mu\nu}\Big\{\bar{g}^{\alpha \mu} D^{\nu}\bar{F}_{\rho \sigma}+ \bar{{g}^{\mu}}_{[\sigma}D_{\rho]}\bar{F}^{\alpha \nu}\Big\}a_{\alpha}
 \\
 &+\sqrt{2}a_{\mu}\Big\{\bar{g}^{\mu \alpha} D^{\beta} \bar{F}_{\rho \sigma}+\bar{{g}^{\alpha}}_{[\sigma}D_{\rho]}\bar{F}^{\mu \beta} \Big\}\Phi_{\alpha \beta}.
 \end{split}
\end{align}
This form of $E$ and $\Omega_{\alpha \beta}$, obtained in \cref{bosonic E,Bosonic Omega} from the field redefinition approach of Seeley-DeWitt expansion \cite{Charles:2015nn} is simpler and shorter as compared to the $E$ and $\Omega_{\alpha \beta}$ evaluated via the general approach of Seeley-Dewitt expansion in \cite{Karan:2019gyn}. So, this in turn also eases the computation of traces of $E$, $E^2$ and $\Omega_{\alpha \beta} \Omega^{\alpha \beta}$, particularly $E^2$ and $\Omega_{\alpha \beta} \Omega^{\alpha \beta}$ which takes a lot of time and effort in the general approach. We get the following traces:
\begin{align}\label{boson trace}
\begin{split}
\text{tr} \thickspace (\mathbb{I})&=9+4+1=14,
\\
\text{tr} \thickspace (E)&=6 \bar{F}^{\mu \nu} \bar{F}_{\mu \nu},
\\
\text{tr} \thickspace (E^2)&=3R^{\mu\nu\theta\phi}R_{\mu\nu\theta\phi}-7R^{\mu\nu}R_{\mu\nu}+9(\bar{F}^{\mu \nu}\bar{F}_{\mu \nu})^2+3R_{\mu\nu\theta\phi} \bar{F}^{\mu\nu}\bar{F}^{\theta \phi},
\\
\text{tr}\thickspace (\Omega^{\alpha \beta}\Omega_{\alpha \beta})&=-7R^{\mu\nu\theta\phi}R_{\mu\nu\theta\phi}+56R^{\mu\nu}R_{\mu\nu}-18R_{\mu\nu\theta\phi}\bar{F}^{\mu\nu}\bar{F}^{\theta \phi}-54 (\bar{F}^{\mu \nu} \bar{F}_{\mu \nu})^2.
\end{split}
\end{align}
So, the Seeley-DeWitt coefficients for the bosonic sector from the gauge fixed action of \say{non-minimal} $\mathcal{N}=1$, $d=4$ EMSGT can be obtained by using \cref{boson trace,a 2n} and are given as,
\begin{align}\label{boson a_{2n}}
\begin{split}
&(4\pi)^2 a_{0}^{\text{EM}}(x)=14,
\\
&(4\pi)^2 a_{2}^{\text{EM}}(x)=6 \bar{F}^{\mu \nu}\bar{F}_{\mu \nu},
\\
&(4\pi)^2 a_{4}^{\text{EM}}(x)=\frac{1}{180}\big(179R^{\mu\nu\theta\phi}R_{\mu\nu\theta\phi}+196 R^{\mu\nu}R_{\mu\nu}\big).
\end{split}
\end{align}
The above results match perfectly with the Seeley-DeWitt coefficients in \cite{Bhattacharyya:2012ss,Karan:2019gyn} computed through the general approach of Seeley-DeWitt expansion.

During the gauge fixing process of the Einstein-Maxwell action \eqref{non fluctuated bosonic action}, there arise the ghost fields. The action describing the ghost fields is:
\begin{equation}\label{bosonic ghost action}
	\mathcal{S}_{\text{ghost},b}=\frac{1}{2}\int d^4x \sqrt{\bar{g}}\lbrace2b_{\mu}(\bar{g}^{\mu\nu}\Box+R^{\mu\nu})c_{\nu}+2b\Box c-4b\bar{F}^{\rho \nu}D_{\rho}c_{\nu}  \rbrace,
\end{equation}
where $b_{\mu}$ and $c_{\mu}$ are vector fields that are associated with diffeomorphism ghosts, $b$ and $c$ are scalar ghost fields which are associated with graviphoton in \cref{bosonic ghost action}. The 
Seeley-DeWitt coefficients for this part can be computed following the 
general approach as the action is already in a simplified form. We recapitulate the results from our earlier work \cite{Karan:2019gyn},\footnote{ For details of the computation, one can refer to this reference.}
\begin{align}\label{ghost b a_{2n}}
\begin{split}
&(4 \pi)^2 a_{0}^{\text{ghost},b}(x)=-10,
\\
&(4 \pi)^2 a_{2}^{\text{ghost},b}(x)=0,
\\
&(4 \pi)^2 a_{4}^{\text{ghost},b}(x)=\frac{1}{18}\big(2R^{\mu\nu\theta\phi}R_{\mu\nu\theta\phi}-17 R^{\mu\nu}R_{\mu\nu}\big).
\end{split}
\end{align} 
The net Seeley-DeWitt coefficients for the bosonic sector of \say{non-minimal} $\mathcal{N}=1$, $d=4$ EMSGT will be obtained by summing \cref{ghost b a_{2n},boson a_{2n}}, which yields
\begin{align}\label{a_{2n} B}
\begin{split}
&(4\pi)^{2}a_{0}^{\text{B}}(x)=4,
\\
&(4\pi)^{2} a_{2}^{\text{B}}(x)=6\bar{F}^{\mu\nu}\bar{F}_{\mu\nu},
\\
&(4\pi)^{2}a_{4}^{\text{B}}(x)=\frac{1}{180}\big(199 R^{\mu\nu\theta\phi}R_{\mu\nu\theta\phi}+26R^{\mu\nu}R_{\mu\nu}\big).
\end{split}
\end{align}
\subsection{Seeley-DeWitt coefficients for fermionic sector}\label{fermions}

Here, in this section, we shall compute the first three Seeley-Dewitt coefficients for the fermionic sector of \say{non-minimal} $\mathcal{N}=1$, $d=4$ EMSGT following the field redefinition approach of Seeley-DeWitt expansion \cite{Charles:2015nn}. The fermion sector of this theory is defined with gravitino field $\psi_{\mu}$ and gaugino field $\lambda$, coupled together in a non-minimal way via field strength $F_{\mu\nu}$. The relevant Lagrangian is given in \cref{fermionic action}. 

We begin with the gauge fixed quadratic fluctuated action for the fermionic sector of \say{non-minimal} $\mathcal{N}=1$, $d=4$ EMSGT without ghost \cite{Ferrara,Banerjee:2020wbr}, which is given by
\begin{equation}\label{fermion action}
\mathcal{S}_2=-\frac{1}{2}\int d^4 x \sqrt{ \bar{g}}\tilde{\varphi}_{m}O^{mn}\tilde{\varphi}_{n},
\end{equation} 
\begin{align}\label{fermion operator}
\begin{split}
\tilde{\varphi}_{m}O^{mn}\tilde{\varphi}_{n}=&-\frac{i}{2}\bar{\psi}_{\mu}\gamma^{\nu}\gamma^{\rho}\gamma^{\mu}D_{\rho}\psi_{\nu}+i\bar{\lambda} \gamma^{\rho} D_{\rho} \lambda
\\
&-\frac{i}{2\sqrt{2}}\bar{\psi}_{\mu}\bar{F}_{\alpha \beta}\gamma^{\alpha}\gamma^{\beta}\gamma^{\mu}  \lambda +\frac{i}{2\sqrt{2}}\bar{\lambda}\gamma^{\nu}\gamma^{\alpha}\gamma^{\beta}\bar{F}_{\alpha \beta}\psi_{\nu}.
\end{split}
\end{align}
In order to calibrate \cref{fermion operator}  to a  simplified form, we consider the following gravitino field redefinition,
\begin{equation}\label{field redefinition fermion}
\Psi_{\mu}=\psi_{\mu}-\frac{1}{2}\gamma_{\mu}\gamma^{\nu}\psi_{\nu},
\end{equation} 
The redefined field, $\Psi_{\mu}$, disentangles the computation of Seeley-DeWitt coefficients via the general approach performed in \cite{Banerjee:2020wbr}.
 The action \eqref{fermion action} then takes the form, 
\begin{equation}\label{field redifined action}
=-\frac{1}{2}\int d^4 x \sqrt{\bar{g}}\tilde{\varphi}_{m} O'^{\thinspace mn} \tilde{\varphi}_{n},
\end{equation}
where the operator $O'^{\thinspace mn}$ in Dirac form and is given as
\begin{align}\label{operator O} 
\begin{split}
\tilde{\varphi}_{m} O'^{\thinspace mn} \tilde{\varphi_{n}}=&i\bar{g}^{\mu\nu}\bar{\Psi}_{\mu}\gamma^{\rho}D_{\rho}  \Psi_{\nu} +i \bar{\lambda} \gamma^{\rho}D_{\rho} \lambda
\\
& -\frac{i}{2\sqrt{2}}\bar{\Psi}_{\mu} \bar{F}_{\alpha \beta}\gamma^{\alpha}\gamma^{\beta}\gamma^{\mu}  \lambda +\frac{i}{2\sqrt{2}}\bar{\lambda} \gamma^{\nu}\gamma^{\alpha} \gamma^{\beta} \bar{F}_{\alpha \beta} \Psi_{\nu}.
\end{split}
\end{align}
The field redefinition mainly simplifies the kinetic part of gravitino field contribution from \cref{fermion operator}, which in turn reduces our further computation and trace calculation to a great extent.
The space-time under consideration is even-dimensional having Euclidean signature, which yields the gamma matrices to be Hermitian. So, the form of $O'^{\thinspace mn}$ \eqref{operator O} is Hermitian, Dirac type and  in linear form. For the heat kernel analysis, the form of \cref{operator O} is made Laplacian type, and then the functional determinant is computed following\cite{Peixoto:2001wx,Charles:2015nn,Sen:2011ba}:
\begin{equation}\label{K2}
\Lambda^{mn}={\left(O' \right)^{mp}} {\left(O' \right)_{p}}^{n\dagger}.
\end{equation}
The form of differential operator $\Lambda^{mn}$ obtained after operating \cref{K2} is
\begin{align}\label{fermion Lambda}
\begin{split}
\tilde{\varphi}_{m} \Lambda^{mn} \tilde{\varphi}_{n}=
&-\bar{\Psi}_{\mu}\bigg\lbrace\mathbb{I}_{4}\bar{g}^{\mu\nu} D^{\rho} D_{\rho}+\frac{1}{2}\gamma^{\rho}\gamma^{\sigma} {R^{\mu\nu}}_{\rho \sigma}-\frac{1}{8}\gamma^{\alpha} \gamma^\beta \gamma^\mu \gamma^\nu \gamma^\theta \gamma^\phi\bar{F}_{\alpha \beta} \bar{F}_{\theta \phi}\bigg\rbrace\Psi_{\nu}
\\
& -\bar{\lambda}\bigg\lbrace\mathbb{I}_4D^{\rho} D_{\rho} -\frac{1}{8}\gamma^{\tau}\gamma^{\alpha}\gamma^{\beta}\gamma^{\theta}\gamma^{\phi}\gamma_{\tau} \bar{F}_{\alpha \beta} \bar{F}_{\theta \phi} \bigg\rbrace\lambda
\\
&+\bar{\Psi}_{\mu}\left\{\frac{1}{2\sqrt{2}}\gamma^{\rho}\gamma^{\theta}\gamma^{\phi}\gamma^{\mu}(D_{\rho}\bar{F}_{\theta \phi})\right\}\lambda-\bar{\lambda}\bigg\lbrace\frac{1}{2\sqrt{2}}\gamma^{\rho}\gamma^{\nu}\gamma^{\theta}\gamma^{\phi}(D_{\rho} \bar{F}_{\theta \phi})\bigg\rbrace \Psi_{\nu}
\\
&+\bar{\Psi}_{\mu}\bigg\lbrace\frac{1}{2\sqrt2}(\gamma^{\rho}\gamma^{\theta}\gamma^{\phi}\gamma^{\mu}+\gamma^{\theta}\gamma^{\phi}\gamma^{\mu}\gamma^{\rho})\bar{F}_{\theta \phi}  \bigg\rbrace D_{\rho}\lambda
\\
&-\bar{\lambda}\bigg\lbrace\frac{1}{2\sqrt{2}}(\gamma^{\nu} \gamma^{\theta}\gamma^{\phi}\gamma^{\rho}+\gamma^{\rho}\gamma^{\nu}\gamma^{\theta}\gamma^{\phi}) \bar{F}_{\theta \phi}  \bigg\rbrace D_{\rho}\Psi_{\nu}.
\end{split}
\end{align}
Following the similar steps as of the bosonic sector of this theory in \ref{bosons}, one can determine the form of $P$ and $N^{\rho}$ from $\Lambda$ using \cref{lambda 1,fermion Lambda} as

\begin{align}
\begin{split}\label{P fermion}
\tilde{\varphi}_{m} P^{mn}\tilde{\varphi}_{n}=
&\bar{\Psi}_{\mu}\left\{\frac{1}{2}\gamma^{\rho}\gamma^{\sigma}{R^{\mu\nu}}_{\rho \sigma}-\frac{1}{8}\gamma^{\alpha}\gamma^{\beta}\gamma^{\mu}\gamma^{\nu}\gamma^{\theta}\gamma^{\phi}\bar{F}_{\alpha \beta} \bar{F}_{\theta \phi} \right\}\Psi_{\nu}
\\
&-\bar{\lambda}\left\{\frac{1}{8} \gamma^{\tau}\gamma^{\alpha}\gamma^{\beta}\gamma^{\theta}\gamma^{\phi}\gamma_{\tau}\bar{F}_{\alpha \beta} \bar{F}_{\theta \phi} \right\}\lambda-\bar{\Psi}_{\mu}\bigg\lbrace\frac{1}{2\sqrt{2}}\gamma^{\rho}\gamma^{\theta}\gamma^{\phi}\gamma^{\mu}D_{\rho}\bar{F}_{\theta \phi} \bigg\rbrace\lambda
\\
&
+\bar{\lambda}\bigg\lbrace\frac{1}{2\sqrt{2}}\gamma^{\rho}\gamma^{\nu}\gamma^{\theta}\gamma^{\phi}D_{\rho}\bar{F}_{\theta \phi} \bigg\rbrace\Psi_{\nu},
\end{split}
\\
\begin{split}\label{N fermion}
\tilde{\varphi}_{m}(N^\rho)^{mn}\tilde{\varphi}_{n}=
&-\frac{1}{2\sqrt{2}}\bar{\Psi}_{\mu}\bigg\lbrace\gamma^{\rho}\gamma^{\theta}\gamma^{\phi}\gamma^{\mu}+\gamma^\theta \gamma^\phi \gamma^{\mu}\gamma^{\rho} \bigg\rbrace \bar{F}_{\theta \phi} \lambda
 \\
 &+\frac{1}{2\sqrt{2}}\bar{\lambda}\bigg\lbrace\gamma^{\nu}\gamma^{\theta}\gamma^{\phi}\gamma^{\rho}+\gamma^{\rho}\gamma^{\nu}\gamma^{\theta}\gamma^{\phi}\bigg\rbrace\bar{F}_{\theta \phi}\Psi_{\nu}.
 \end{split}
\end{align}
The expression of $\omega_{\rho}$ can be determined from $N_{\rho}$ using \cref{heat kernel parameters,N fermion}, which on further simplification using gamma matrices properties reduces to the form
\begin{align}\label{field redefined omega}
\begin{split}
\tilde{\varphi}_{m}(\omega^{\rho})^{mn}\tilde{\varphi}_n=
&-\frac{1}{2\sqrt{2}}\bar{\Psi}_{\mu}\bigg\lbrace2\gamma^{\beta}\gamma^{\mu}\bar{F^{\rho}}_{\beta}+\bar{g}^{\rho \mu}\gamma^{\theta}\gamma^{\phi}\bar{F}_{\theta \phi} \bigg\rbrace\lambda
\\
&+\frac{1}{2\sqrt{2}}\bar{\lambda}\bigg\lbrace2\gamma^{\nu}\gamma^{\theta}\bar{F_{\theta}}^{\rho}+\bar{g}^{\rho \nu} \gamma^{\theta}\gamma^{\phi}\bar{F}_{\theta \phi} \bigg\rbrace\Psi_{\nu}.
\end{split}
\end{align}
 The expression of $\mathbb{I}$, $E$ and $\Omega_{\alpha \beta}$ will be given using \cref{fermion Lambda,heat kernel parameters,Omega field strength,P fermion,field redefined omega} as,
 \begin{equation}
 	\begin{split}\label{I fermion}
 		\tilde{\varphi}_{m}\mathbb{I}^{mn}\tilde\varphi_{n}=&\bar{\Psi}_{\mu}\mathbb{I}_{4}\bar{g}^{\mu\nu}\Psi_{\nu}+\bar{\lambda}\mathbb{I}_{4}\lambda,	
 	\end{split}
 \end{equation}
\begin{align}\label{field redefined E}
\begin{split}
\tilde{\varphi}_m E^{mn} \tilde{\varphi}_{n}=
&\bar{\Psi}_{\mu}\bigg\lbrace\frac{1}{2}\gamma^{\rho}\gamma^{\sigma}{R^{\mu\nu}}_{\rho \sigma}-\frac{1}{8} \gamma^{\alpha}\gamma^{\beta}\gamma^{\mu}\gamma^{\nu}\gamma^{\theta}\gamma^{\phi}\bar{F}_{\alpha \beta}\bar{F}_{\theta \phi}-\frac{1}{2}\gamma^{\nu}\gamma^{\theta}{R^{\mu}}_{\theta}+\frac{1}{2}\gamma^{\mu}\gamma^{\theta}{R^{\nu}}_{\theta}
\\
&-\frac{1}{2}\mathbb{I}_{4}\bar{g}^{\mu\nu}\bar{F}^{\theta\phi}\bar{F}_{\theta\phi} +\frac{1}{4}\gamma^{\beta}\gamma^{\mu}\gamma^{\theta}\gamma^{\phi}{\bar{F^{\nu}}}_{\beta} \bar{F}_{\theta \phi}-\frac{1}{4}\gamma^{\alpha}\gamma^{\beta}\gamma^{\nu}\gamma^{\theta}\bar{{F^{\mu}}}_{\theta} \bar{F}_{\alpha \beta}
\\
&+\frac{1}{8}\bar{g}^{\mu\nu}\gamma^{\alpha}\gamma^{\beta}\gamma^{\theta}\gamma^{\phi}\bar{F}_{\alpha \beta} \bar{F}_{\theta \phi}\bigg\rbrace\Psi_{\nu}+\bar{\lambda}\bigg\lbrace\frac{1}{2}\gamma^{\theta}\gamma^{\phi}\gamma^{\alpha}\gamma^{\beta} \bar{F}_{\theta \phi} \bar{F}_{\alpha \beta}\bigg\rbrace \lambda
\\
&-\frac{1}{2\sqrt{2}}\bar{\Psi}_{\mu}\bigg\lbrace\gamma^{\rho}\gamma^{\theta}\gamma^{\phi}\gamma^{\mu}D_{\rho} \bar{F}_{\theta \phi}-\gamma^{\theta}\gamma^{\phi} D^{\mu} \bar{F}_{\theta \phi} \bigg\rbrace\lambda
\\
&+\frac{1}{2\sqrt{2}}\bar{\lambda}\bigg\lbrace\gamma^{\rho}\gamma^{\nu}\gamma^{\theta}\gamma^{\phi}D_{\rho}\bar{F}_{\theta \phi}-\gamma^{\theta}\gamma^{\phi}D^{\nu}\bar{F}_{\theta \phi} \bigg\rbrace\Psi_{\nu},
\end{split}
\end{align}
\begin{align}\label{field redefine Omega}
\begin{split}
\tilde{\varphi}_{m}(\Omega_{\alpha \beta})^{mn}\tilde{\varphi_{n}}=&\tilde{\varphi}_{m}[D_{\alpha}, D_{\beta}]^{mn}\tilde{\varphi}_{n}+\tilde{\varphi}_{m}(D_{[\alpha}\omega_{\beta]})^{mn}\tilde{\varphi}_{n}+\tilde{\varphi}_{m}[\omega_{\alpha},\omega_{\beta}]^{mn}\tilde{\varphi}_{n}
\\
=&\bar{\Psi}_{\mu}\bigg\lbrace\mathbb{I}_4{R^{\mu\nu}}_{\alpha \beta} +\frac{1}{4}\bar{g}^{\mu\nu}R_{\alpha \beta \xi \kappa}\gamma^{\xi}\gamma^{\kappa}+\Big(-\frac{1}{2}\gamma^{\rho}\gamma^{\mu}\gamma^{\nu}\gamma^{\sigma}\bar{F}_{\rho \alpha}\bar{F}_{\beta \sigma}
\\
&+\frac{1}{4}\bar{g_{\beta}}^{\nu}\gamma^{\rho}\gamma^{\mu}\gamma^{\eta}\gamma^{\kappa}\bar{F}_{\rho \alpha}\bar{F}_{\eta \kappa}+\frac{1}{4}\bar{g_{\alpha}}^ {\mu}\gamma^{\theta}\gamma^{\phi}\gamma^{\nu}\gamma^{\rho}\bar{F}_{\theta \phi}\bar{F}_{\beta \rho}
\\
&-\frac{1}{8}\bar{g_{\alpha}} ^{\mu}\bar{g_{\beta}}^{\nu}\gamma^{\theta}\gamma^{\phi}\gamma^{\eta}\gamma^{\kappa}\bar{F}_{\theta \phi}\bar{F}_{\eta \kappa}-(\alpha\leftrightarrow\beta)\Big)\bigg\rbrace\Psi_{\nu}
\\
& +\bar{\lambda}\bigg\lbrace\frac{1}{4}\gamma^\theta \gamma^\phi R_{\alpha\beta\theta\phi}+\Big(-\frac{1}{2}\gamma^{\mu}\gamma^{\rho}\gamma^{\eta}\gamma_{\mu}\bar{F}_{\alpha \rho}\bar{F}_{\eta \beta}+\frac{1}{4}\gamma_{\beta}\gamma^{\rho}\gamma^{\xi}\gamma^{\kappa} \bar{F}_{\alpha \rho}\bar{F}_{\xi\kappa}
\\
&+\frac{1}{4}\gamma^{\theta}\gamma^{\phi}\gamma^{\eta}\gamma_{\alpha}\bar{F}_{\theta \phi} \bar{F}_{\eta \beta}-\frac{1}{8}\bar{g}_{\alpha \beta}\gamma^{\theta}\gamma^{\phi} \gamma^{\xi} \gamma^{\kappa}\bar{F}_{\theta \phi}\bar{F}_{\xi\kappa}-(\alpha\leftrightarrow\beta)\Big) \bigg\rbrace\lambda
\\
&-\frac{1}{2\sqrt{2}}\bar{\Psi}_{\mu}\bigg\lbrace 2\gamma^\rho \gamma^\mu D_{\alpha}\bar{F}_{\beta \rho}+\bar{g_{\beta}}^{\mu}\gamma^\theta \gamma^\phi D_{\alpha}\bar{F}_{\theta \phi}-(\alpha\leftrightarrow\beta)\bigg\rbrace\lambda
\\
&+\frac{1}{2\sqrt{2}}\bar{\lambda}\bigg\lbrace2\gamma^{\nu}\gamma^{\theta}D_{\alpha}\bar{F}_{\theta\beta }+\bar{g_{\beta}}^{\nu}\gamma^{\theta}\gamma^{\phi}D_{\alpha}\bar{F}_{\theta \phi}-(\alpha\leftrightarrow\beta) \bigg\rbrace\Psi_{\nu}.
\end{split}
\end{align}
The expressions of $E$ and $\Omega_{\alpha \beta}$ in \cref{field redefined E,field redefine Omega} are simpler than the form of the same obtained in \cite{Banerjee:2020wbr} computed following the general approach. This makes further trace computations fast and straightforward. The traces are calculated as:
\begin{align}\label{trace fermion}
\begin{split}
\text{tr}\thickspace (\mathbb{I})&=16+4=20,
\\
\text{tr} \thickspace (E)&=-8 \bar{F}^{\mu \nu}\bar{F}_{\mu\nu},
\\
\text{tr}\thickspace (E^2)&=10(\bar{F}^{\mu\nu}\bar{F}_{\mu\nu})^2+3R^{\mu\nu}R_{\mu\nu}+2R^{\mu\nu\theta\phi}R_{\mu\nu\theta\phi}-2R_{\alpha\beta\theta\phi}\bar{F}^{\alpha\beta}\bar{F}^{\theta\phi},
\\\text{tr} \thickspace (\Omega^{\alpha \beta}\Omega_{\alpha\beta})&=-\frac{13}{2}R^{\mu\nu\theta\phi}R_{\mu\nu\theta\phi}+12R_{\alpha\beta\theta\phi}\bar{F}^{\alpha\beta}\bar{F}^{\theta\phi}-6R^{\mu\nu}R_{\mu\nu}
\\
&\qquad-60(\bar{F}^{\mu\nu}\bar{F}_{\mu\nu})^2.
\end{split}
\end{align}
Using the trace results \eqref{trace fermion} in \eqref{a 2n}, the required Seeley-DeWitt coefficients for the fermionic sector of gauge-fixed \say{non-minimal} $\mathcal{N}=1$, $d=4$ EMSGT are 
\begin{align}\label{fermion coefficients}
\begin{split}
& (4\pi)^2 a_{0}^{f}(x)=-10,
\\
& (4\pi)^2 a_{2}^{f}(x)=4 \bar{F}^{\mu\nu} \bar{F}_{\mu\nu},
\\
& (4\pi)^2a_{4}^{f}(x)=-\frac{1}{144}\left(41 R^{\mu\nu\theta\phi} R_{\mu\nu\theta\phi}+64R^{\mu\nu}R_{\mu\nu} \right).
\end{split}
\end{align}
Considering the fermion spin-statistics 
 and Majorana degree of freedom, a factor of -1/2 is inserted manually in \cref{fermion coefficients}. The results \eqref{fermion coefficients} are consistent with the results of the same in our earlier work \cite{Banerjee:2020wbr}.
 
 Now, we work out on determination of Seeley-DeWitt coefficients for the ghost part in the fermionic sector of \say{non-minimal} $\mathcal{N}=1$, $d=4$ EMSGT. The Lagrangian for this ghost sector is given as
\begin{equation}\label{fermion ghost}
\mathcal{L}_{\text{ghost}}=\bar{\tilde{b}}\gamma^{\rho}D_{\rho}\tilde{c}+\bar{\tilde{e}}\gamma^{\rho}D_{\rho}\tilde{e}.
\end{equation}
$\tilde{b}$, $\tilde{c}$ and $\tilde{e}$ are bosonic ghosts that are composed of three minimally coupled Majorana fermions obeying spin 1/2 statistics. The computation of Seeley-DeWitt coefficients for this case is very simple and upfront, which is given by (-3) times of Seeley-DeWitt coefficients for free Majorana spin 1/2 field.  The results are \cite{Banerjee:2020wbr,Karan:2017txu}:
\begin{align}\label{ghost contribution fermion}
\begin{split}
&(4\pi)^2 a_{0}^{\text{ghost},f}(x)=6,
\\
&(4\pi)^2 a_{2}^{\text{ghost},f}(x)=0,
\\
&(4\pi)^2 a_{4}^{\text{ghost},f}(x)=-\frac{1}{240}\left(7 R^{\mu\nu\theta \phi} R_{\mu\nu\theta\phi} + 8R_{\mu\nu}R^{\mu\nu} \right).
\end{split}
\end{align}
Net fermionic Seeley-DeWitt coefficients can be obtained by summing \cref{fermion coefficients,ghost contribution fermion} for \say{non-minimal} $\mathcal{N}=1$, $d=4$ EMSGT, which is given by
\begin{align}\label{a_{2n} F}
\begin{split}
&(4\pi)^2 a_{0}^{\text{F}}(x)=-4,
\\
&(4\pi)^2 a_{2}^{\text{F}}(x)=4\bar{F}^{\mu\nu}\bar{F}_{\mu\nu},
\\
&(4\pi)^2 a_{4}^{\text{F}}(x)=-\frac{1}{360}\left(113 R^{\mu\nu\theta\phi} R_{\mu\nu\theta\phi}+172 R^{\mu\nu}R_{\mu\nu} \right).
\end{split}
\end{align} 
\subsection{Total Seeley-DeWitt coefficients}\label{total seeley-Witt coefficients}
Once we have determined the bosonic and fermionic Seeley-DeWitt coefficients for \say{non-minimal} $\mathcal{N}=1$, $d=4$ EMSGT, the total Seeley-DeWitt coefficients are obtained after adding both the sectors, \eqref{a_{2n} B} and \eqref{a_{2n} F}. The results are
\begin{align}\label{total a2n}
\begin{split}
&(4\pi)^2 a_{0}^{\mathcal{N}=1}(x)=0,
\\
&(4\pi)^{2} a_{2}^{\mathcal{N}=1}(x)=10 \bar{F}^{\mu\nu}\bar{F}_{\mu\nu},
\\
&
(4\pi)^2 a_{4}^{\mathcal{N}=1}(x)=\frac{1}{24}\left(19 R^{\mu\nu\theta\phi}R_{\mu\nu\theta\phi}-8R^{\mu\nu}R_{\mu\nu} \right).
\end{split}
\end{align}
The vanishing of $a_0$ in \eqref{total a2n} clearly shows the presence of an equal number of bosonic and fermionic degrees of freedom in the \say{non-minimal} $\mathcal{N}=1$, $d=4$ EMSGT. 
Note that $a_4$ obtained in \eqref{total a2n} only depends upon the background metric and thus preserves rotational invariance under electric and magnetic duality. Among the Seeley-DeWitt coefficients in \eqref{total a2n}, we are mainly focused on $a_4$ because it determines the logarithmic divergence of the black hole entropy in large charge limit.
This $a_4$ can also be represented in terms of the square of Weyl tensor $W_{\mu\nu\rho\sigma}$ and Gauss-Bonnet term (Euler density) $E_{4}$ as
\begin{equation}\label{a4 W E}
(4 \pi)^2 a_{4}(x)=c W^{\mu\nu\rho\sigma}W_{\mu\nu\rho\sigma}- a E_{4},
\end{equation}
where $c$ and $a$ are constants, which depend upon the field content and couplings among them in the theory . $W^{\mu\nu\rho\sigma}W_{\mu\nu\rho\sigma}$ and $E_{4}$ is defined as
\begin{equation}\label{W}
W^{\mu\nu\rho\sigma}W_{\mu\nu\rho\sigma}=R^{\mu\nu\rho\sigma}R_{\mu\nu\rho\sigma}-2R^{\mu\nu}R_{\mu\nu}+\frac{1}{3}R^2
\end{equation}
and
\begin{equation}\label{E}
E_{4}=R^{\mu\nu\rho\sigma}R_{\mu\nu\rho\sigma}-4R^{\mu\nu}R_{\mu\nu}+R^2.
\end{equation}
Using \cref{total a2n,a4 W E,W,E}, $a_{4}$ may also be represented as\footnote{We have set $R=0$ in the definition of \cref{W,E}.}
\begin{equation}\label{a4}
(4 \pi)^2 a_{4}^{\mathcal{N}=1}(x)=\frac{17}{12}W^{\mu\nu\rho\sigma}W_{\mu\nu\rho\sigma}-\frac{5}{8}E_{4}.
\end{equation}
This representation of $a_4$ will be helpful in our computation of logarithmic corrections to the entropy of non-extremal black holes in the next section.


\section{Logarithmic corrections to non-extremal black holes in ``non-minimal" $\mathcal{N}=1$, $d=4$ EMSGT}\label{section 5}
In this section, we first review the formalism from \cite{Sen:2012dw} and \cite{Charles:2018yey,Karan:2021teq}\footnote{  Ref. \cite{Charles:2018yey,Karan:2021teq} presents a review of the formalism of \cite{Sen:2012dw}.} to compute the logarithmic corrections to the entropy of generic non-extremal black holes. We particularly focus on Kerr-Newman family of black holes.
Then, we determine the corresponding logarithmic corrections to these non-extremal black holes using the Seeley-DeWitt coefficient $a_{4}$ presented in \eqref{total a2n}.
\subsection{General framework} \label{GenFrm}
Here, we present the general framework of Euclidean gravity approach for the determination of logarithmic corrections to the entropy of generic non-extremal black holes in four dimensions following \cite{Sen:2012dw,Charles:2018yey}. 

We begin with grand canonical partition function $\mathcal{Z}(\beta,\vec{\omega},\vec{\mu})$\footnote{$\beta$, $\vec{\omega}$ and $\vec{\mu}$ are black hole potentials.}, which is defined as Euclidean path integral of the action of a theory \eqref{partition function}. This integral is evaluated using saddle point approximation, followed by expansion \eqref{field fluctuations}. The classical saddle points in \cref{field fluctuations} describes the black holes that are in equilibrium with dilute gas of thermal non-interacting particles (Hawking particle) in the theory. The first task is to separate the contribution to the partition function associated with the thermal gas and extract the partition function associated only with black holes. For this purpose, first, we will evaluate the partition function associated with thermal gas.   

We consider the same Euclidean spacetime where the time coordinate has periodicity $\beta$, the spatial coordinates are fixed to length $L$, and the black hole is confined in a box of length $L$. The kinetic operator acts over the thermal gas and yields the eigenvalues:
\begin{equation}
	\frac{4 \pi^2 n^2}{\beta^2}+\vec{k}^2,
\end{equation}
$n$ defines the momentum along compact time direction and $\vec{k}$ denotes the spatial momentum. The density of states associated with this momentum in a large volume limit is given by
\begin{equation}
d\mu=\frac{V}{(2 \pi)^3}d^3{k},
\end{equation}
$V=L^3$ is the volume of the box.
So, the one-loop quantum effective action \eqref{one loop effective action} is thus given by
\begin{equation}\label{effective action thermal gas}
	W=-\frac{1}{2}\int_{\epsilon}^{\infty}\frac{ds}{s} \sum_{n=-\infty}^{\infty}\int d\mu \exp\Big(-\frac{4 \pi^2 n^2 s}{\beta^2}-s\vec{k}^2\Big).
\end{equation}	
The above integrand is divergent for all values of $n$ in limit $s \to 0$. In order to solve the above integration \eqref{effective action thermal gas}, we use the following identity:
\begin{equation}\label{identity}
	\sum_{n=-\infty}^{\infty}\exp\Big(-\frac{4\pi^2 n^2 s}{\beta^2}\Big)=\frac{\beta}{\sqrt{4 \pi s}}\sum_{n=-\infty}^{\infty}\exp \Big(-\frac{n^2 \beta^2}{4s} \Big).
\end{equation}
Using \cref{identity,effective action thermal gas}, we get
\begin{equation}\label{effective action thermal gas 2}
	W=-\frac{1}{2}\int_{\epsilon}^{\infty}\frac{ds}{s} \frac{\beta}{\sqrt{4\pi s}}\sum_{n=-\infty}^{\infty}\int d\mu \exp \Big(-\frac{n^2 \beta^2}{4s}-s\vec{k}^2 \Big).
\end{equation}
The integral \eqref{effective action thermal gas 2} leads to the one-loop effective action for thermal gas \cite{Charles:2018yey}:

\begin{equation}\label{effective action thermal gas 1}
	W=-\frac{V \beta}{64 \pi^2 \epsilon^2}-\frac{\pi^2 V}{90 \beta^{3}}.
\end{equation}
 The first term is ultraviolet divergent at $\epsilon \to 0$. The contribution \eqref{effective action thermal gas 1} belongs to the bulk part, which is proportional to the volume of the box $L^3$. Again, there will be boundary contributions, which will give the subleading corrections to the one-loop effective action \eqref{effective action thermal gas 1}. Now, we will consider this boundary effect and find the required one-loop partition function. It will involve the contributions which are lower in the power of $L$. The density of states will be modified to
\begin{equation}\label{density of states boundary}
	d\mu=\frac{V}{(2 \pi)^3}d^3k+\mathcal{O}(L^2d^2k),
\end{equation}  
 where the second term in \eqref{density of states boundary} is the boundary contribution. By this modification \eqref{density of states boundary} in the density of states and dropping the divergent part, the one-loop effective action \eqref{effective action thermal gas 2} will also be modified. 
  The one-loop effective action, including the boundary part, thus will be given as
 \begin{equation}\label{effective action thermal gas 3}
 	W=-\frac{\pi^2 V}{90 \beta^3}+\mathcal{O}\Big(\frac{L^2}{\beta^2} \Big).
 \end{equation}
In case of $L \gg \beta$ (large volume limit), the dominant contribution will come from the bulk part in \eqref{effective action thermal gas 3}. This one-loop effective action is associated with thermal gas that is propagating in Euclidean flat spacetime. Extending this thermal gas to a black hole background having potentials $\beta$, $\vec{\omega}$ and $\vec{\mu}$, where $\beta$ is the inverse temperature, $\vec{\omega}$ is the angular velocity and $\vec{\mu}$ is the chemical potential of black hole having radius $a$. The thermal gas is in equilibrium with the black hole. Then, the one-loop effective action for thermal gas in large volume limit in this black hole background is thus given as 
\begin{equation}\label{effective action final}
	W=L^3 f(\beta,\vec{\omega}, \vec{\mu})+\mathcal{O}\Big( \frac{L^2}{\beta^2}\Big),
\end{equation}
where $f$ is a function that scales as follows:
\begin{equation}\label{function}
	f(\lambda \beta, \lambda \vec{\omega}, \lambda \vec{\mu})=\lambda^{-3} f(\beta,\vec{\omega}, \vec{\mu}).
\end{equation}
   Here, we consider a black hole with solution $(\bar{g}_{\mu\nu},\bar{A}_{\mu})$ having radius $a$ and confined in a box of volume $V=L^3$ along with thermal gas. The contribution by the thermal gas to the effective action is given by \eqref{effective action final}. In order to subtract off the contribution due to thermal gas \eqref{effective action final}, and extract the contribution that only corresponds to black hole, we consider another black hole having radius $a'$ and related to the previous black hole with solution $(\xi^2\bar{g}_{\mu\nu},\xi \bar{A}_{\mu})$, where $\xi=a'/a$. This black hole is also confined in an identical box having a length $L'=L \thickspace a'/a$. The boundary conditions on the fields to the second black hole are also related to the first one by scale transformation $\xi$. Then the one-loop effective action due to thermal gas by this second box using \cref{effective action final,function} will be given as
  \begin{equation}\label{effective action system 2}
  	W=\Big(L\frac{a'}{a}\Big)^3 f\Big(\frac{a'}{a}\beta, \vec{\omega},\frac{a'}{a}\vec{\mu}\Big)+\mathcal{O}\Big(\frac{L^2}{\beta^2} \Big)=L^{3}f(\beta, \vec{\omega}, \vec{\mu})+\mathcal{O}\Big(\frac{L^2}{\beta^2} \Big).
  \end{equation} 
So, from \cref{effective action final,effective action system 2}, we infer that the thermal contributions due to both black holes are same. If we subtract out the one-loop effective actions of both the black holes confined in two different systems, then the contribution due to thermal gas along with spurious boundary terms will be subtracted out. We will then only be left with the difference between the one-loop effective action associated only for black holes. Say $\Delta W$ is the difference between the non-zero mode contribution to the one-loop effective actions due to both the black holes, then from \cref{one loop effective action} we have
\begin{equation}\label{difference effective action}
\Delta W=-\frac{\chi}{2}\int_{\epsilon}^{\infty} \frac{ds}{s}\sideset{}{'}\sum_{\substack{i\\ \lambda_{i},\lambda^{'}_{i}\ne0}}(e^{-s \lambda_i}-e^{-s\lambda^{'}_i}),
\end{equation}
where $\lambda_{i}$ and $\lambda'_{i}$ are the eigenvalues of kinetic operator associated with black holes having radius $a$ and $a'$ respectively and $\sideset{}{'}\sum$ defines the summation over non-zero values of $\lambda_i$, $\lambda_{i}^{'}$ in effective action \eqref{difference effective action}. Both the eigenvalues of different systems are related as
\begin{equation}\label{relation}
\lambda'_{i}=\lambda_{i}\Big(\frac{a}{a'}\Big)^2.
\end{equation}
Using \cref{relation,difference effective action}, we get
\begin{align}\label{difference effective action 2}
	\begin{split}
\Delta W&=-\frac{\chi}{2}\Big\lbrace\int_{\epsilon}^{\infty} \frac{ds}{s}\sideset{}{'}\sum_{\substack{i \\ \lambda_{i\ne0}}} e^{-s \lambda_i}-\int_{\epsilon}^{\infty}\frac{ds}{s}\sideset{}{'}\sum_{\substack{i\\ \lambda_{i\ne0}}} e^{-s\lambda_i  (a/a')^2}\Big\rbrace
\\
&=-\frac{\chi}{2}\int_{\epsilon}^{\epsilon'}\frac{ds}{s}\sideset{}{'}\sum_{\substack{i\\ \lambda_{i\ne0}}} e^{-s \lambda_{i}}. 
\end{split}
\end{align}
In \cref{difference effective action 2}, we rescale $s (a/a')^2 \to s$ and the UV limit as $\epsilon'=\epsilon(a/a')^2 $. We also get rid of infrared divergences in the above integrand by elimination of the infinity limit. In the integration \eqref{difference effective action 2}, the dominant contribution comes when $s/a^2$ ranges between $\epsilon/a^2$ and $\epsilon/a'^2$. Here $a$ and $a'$ are large compared to UV cut-off $\sqrt{\epsilon}$. So $s/a^2$ remains small over entire range. It fulfills the criteria for small perturbative expansion in proper time, so using \cref{expansion,ds1,difference effective action 2}, the non zero contribution to one-loop effective action:
\begin{equation}\label{effective action and heat kernel}
	\Delta W =-\frac{\chi}{2} \int_{\epsilon}^{\epsilon'}\frac{ds}{s}\int d^4x\sqrt{\bar{g}}\Big(\frac{1}{s^2}a_{0}+\frac{1}{s}a_{2}+a_{4}+....-K^{\text{zm}}(x,x;0)\Big).
\end{equation}    
The logarithmic correction term will arise from $s$ independent term, i.e., $a_4$ in the integrand of \cref{effective action and heat kernel}. So, the above integration \eqref{effective action and heat kernel} yield the one-loop effective action:
\begin{equation}\label{effective action and local-zero mode}
 W \simeq-\frac{\chi}{2}\Big\lbrace \int d^4 x \sqrt{\bar{g}} a_4 +...  -M\Big\rbrace \ln a^2,
\end{equation}
where 
\begin{equation}\label{M}
	M=\int d^4 x \sqrt{\bar{g}} K^{\text{zm}}(x,x;0).
\end{equation}
 $M$ is the number of zero modes associated with field fluctuations in \cref{M}.  

So we started with grand canonical partition function \eqref{partition function} where $\beta$, $\vec{\omega}$ and $\vec{\mu}$ are fixed as well as scale linearly with the size of black hole. We evaluated the one-loop effective action for arbitrary fluctuated fields propagating in black hole background.
 The partition function and effective action are related as
\begin{equation}\label{partition function and action}
\ln \mathcal{Z}(\beta, \vec{\omega},\vec{\mu})=-\mathcal{S_{\text{cl}}}-W,
\end{equation}
 where $\mathcal{S}_{\text{cl}}$ is the classical action. In order to compute the logarithmic corrections part to the entropy of non-extremal black holes, we need to extend the ensemble to microcanonical where mass $(M)$, momentum $(\vec{P})$, angular momentum $(\vec{J})$ and charge $(\vec{Q})$ are fixed. Each microstate of the black hole is associated with relativistic energy $(E)$ as
 \begin{equation}\label{Energy}
 	E=M+\frac{\vec{P}^2}{2M}.
 \end{equation}
 The black hole entropy $S_{BH}$ is the number of states present in the microcanonical ensemble. 
  The microstate degeneracy $(\Omega)$ and entropy $(S_{\text{BH}})$ is related as
 \begin{equation}
 	\Omega(M,\vec{P},\vec{J},\vec{Q})=e^{S_{\text{BH}}(M,\vec{J},\vec{Q})}.
 \end{equation}
   Then the grand canonical partition function can be expressed as sum of all black hole microstate, which is further related to black hole entropy as \cite{Gibbons:1976ue}
   \begin{align}\label{partition function and entropy}
   	\begin{split}
   	\mathcal{Z}(\beta,\vec{\omega},\vec{\mu})&=\sum_{M \vec{P} \vec{J}\vec{Q}}\Omega(M,\vec{P},\vec{J},\vec{Q})
   	\\
   	&=\sum_{M \vec{P} \vec{J} \vec{Q}}e^{S_{\text{BH}}(M,\vec{J},\vec{Q})-\beta E -\vec{\omega}.\vec{J}-\vec{\mu}.\vec{Q}}.
   	\end{split}
   \end{align}
 Using \cref{partition function and action,partition function and entropy,Energy}, the entropy of black hole is defined as
\begin{equation}\label{entropy and action}
	S_{\text{BH}}(M,\vec{J},\vec{Q})=-\mathcal{S}_{\text{cl}}+\beta M+\vec{\omega}.\vec{J}+\vec{\mu}.\vec{Q}-W.
\end{equation}
  The sum over $\vec{P}$ in \cref{entropy and action} has been done implicitly. Now, from \cref{entropy and action}, the computation of one-loop effective action turned into the determination of entropy of a black hole. The first four terms in \cref{entropy and action} belong to Bekenstein-Hawking entropy\cite{hawking}. From \cref{effective action and local-zero mode,entropy and action}, the logarithmic correction to the entropy of a generic non-extremal black hole is given by \cite{Charles:2015nn,Castro:2018hsc}
\begin{align}\label{entropy relation}
	\begin{split}
	\Delta S_{\text{BH}}&=-W
	\\
	&=\Delta S_{\text{local}}+\Delta S_{\text{zm}},
	\end{split}
\end{align} 
where
\begin{equation}\label{local mode}
	\Delta S_{\text{local}}=\frac{\chi}{2}\int d^4 x \sqrt{\bar{g}} a_{4}(x) \ln A_{H},
\end{equation}
and 
\begin{align}\label{zero mode contribution}
	\begin{split}
	\Delta S_{\text{zm}}
	&=\frac{\chi}{2}\Big\{  \sum_{r}(Y_{r}-1)M^r\Big\} \ln A_{H}.
	\end{split}
\end{align}
$A_{H}$ is the horizon area of the black hole in \cref{local mode,zero mode contribution}, which is proportional to $a^2$. $S_{\text{zm}}$ \eqref{zero mode contribution} corresponds to zero mode contributions to the logarithmic corrections. Here $Y$ is the scaling dimension of different field fluctuations present in a theory and $M$ is defined in \eqref{M}. Both parameters possess specific values for different types of fields present in the theory. In \eqref{zero mode contribution}, an additional contribution associated with $Y$ arises due to the fact that the zero modes $(\lambda_i=0)$ are not fully computed by effective action \eqref{effective action and local-zero mode}. To get fully corrected zero mode contribution, the integration over fields is replaced by integration over the zero mode deformation. The jacobian of changing these variables gives $a^{Y_{r}}$ per zero mode. So for $M$ zero modes, we get this additional factor $a^{Y_{r}M^{r}}$, including which we get the corrected zero mode contribution in \eqref{zero mode contribution}. The zero mode contributions \eqref{zero mode contribution} have been computed in various literature \cite{Banerjee:2011jp,Sen:2012rr,Sen:2011ba,Sen:2012dw}, which can be presented compactly as \cite{Charles:2015nn}
\begin{equation}\label{zero mode contribution 2}
\Delta S_{\text{zm}}=\frac{1}{2}\bigg\lbrace-(3+K)+2N_{\text{SUSY}}+3\delta \bigg\rbrace \ln A_{H},
\end{equation}
 where $K$ is the number of rotational isometries, which is defined as 3 for non-rotating black holes and 1 for rotating ones. $N_{\text{SUSY}}$ is the number of preserved supercharges in the supersymmetric theory. The factor 3$\delta$ factor arises from finite IR volume integration in non-extremal black holes, so $\delta$ is set 1 for non-extremal black holes and 0 for extremal black holes.
 
 It is important to note that the structure of relation \eqref{entropy relation} is very similar while we evaluate the same for extremal black holes \cite{Bhattacharyya:2012ss,Karan:2019gyn,Banerjee:2020wbr,Karan:2020njm,Banerjee:2010qc,Banerjee:2011jp,Sen:2012rr,Sen:2011ba,Keeler:2014bra,Karan:2021teq}. Here, the path integral in \eqref{entropy relation} is evaluated over the whole spacetime of black hole background for non-extremal black holes \cite{Sen:2012dw,Charles:2015nn,Castro:2018hsc}. However, in case of extremal black holes, we perform the integration over near horizon geometry using quantum entropy function formalism \cite{Sen:2008yk,Sen:2009vz,Sen:2008vm}, which is not applicable for non-extremal black holes.

\subsection{Logarithmic corrections to non-extremal Kerr-Newman family of black holes in ``non-minimal" $\mathcal{N}=1$, $d=4$ EMSGT}\label{Application of Seeley-DeWitt coefficients in Logarithmic correction of non extremal Black holes}
An $\mathcal{N}=1$, $d=4$ EMSGT can have black hole solutions, which include Kerr-Newman, Kerr, Reissner-Nordstr\"{o}m and Schwarzschild black holes. In this section, our aim is to compute the logarithmic corrections to the entropy of above black holes in the non-extremal limit following the approach discussed in \cref{GenFrm}.

The Kerr-Newman metric defined with mass $M$, charge $Q$ and angular momentum $J$  is given by
\begin{align}\label{metric}
\begin{split}
ds^2=&-\frac{r^2+b^2\cos^2 \psi-2Mr+Q^2}{r^2+b^2\cos^2\psi}dt^2+\frac{r^2+b^2\cos^2 \psi}{r^2+b^2-2Mr+Q^2}dr^2
\\
&+\bigg(\frac{(r^2+b^2\cos^2 \psi)(r^2+b^2)+(2Mr-Q^2)b^2\sin^2 \psi}{r^2+b^2\cos^2\psi}\bigg)\sin^2\psi d\phi^2
\\
&+(r^2+b^2\cos^2 \psi)d\psi^2+\frac{2(Q^2-2Mr)b}{r^2+b^2\cos^2\psi}\sin^2\psi dt d\phi,
\end{split}
\end{align}
where 
\begin{align}\label{parameter}
 b=J/M \enspace\text{and}\enspace r_{H}=M+\sqrt{M^2-Q^2-b^2}.
\end{align}
 The event horizon radius $r_{H}$ in \cref{parameter} is associated with condition $M^2\gtreqless Q^2+b^2$, where  $M^2= Q^2+b^2$ corresponds to extremal black hole solutions and  $M^2> Q^2+b^2$ corresponds to the non-extremal black hole solutions. In contrast, the black hole solutions corresponding to $M^2< Q^2+b^2$ is physically unacceptable because it does not possess an event horizon and exhibit naked singularity. Here, we are interested in the non-extremal solution associated with metric \eqref{metric}. The inverse temperature $\beta$ scales as length scale for the case of non-extremal black hole. However, in the case of extremal black holes,  $\beta \to \infty$ leading to $T \to 0$. The classical entropy is given as
\begin{equation}\label{scl}
S_{\text{cl}}=\frac{A_{H}}{4G_{N}}=16\pi^2 (2M^2-Q^2+2M\sqrt{M^2-b^2-Q^2}),
\end{equation}
with
\begin{align}\label{beta}
	\begin{split}
	 \beta&=\frac{\partial S_{BH}}{\partial M}=\frac{32\pi^2 }{\sqrt{M^2-b^2-Q^2}}\bigg\lbrace2M^2-Q^2+2M\sqrt{M^2-b^2-Q^2}\bigg\rbrace,
	 \end{split}
\\
	\begin{split}\label{omega}
		\omega&=\frac{\partial S_{BH}}{\partial J}=-\frac{32\pi^2 b}{\sqrt{M^2-b^2-Q^2}},
	\end{split}
\\
	\begin{split}\label{mu}
		\mu&=\frac{\partial S_{BH}}{\partial Q}=-\frac{32\pi^2 Q}{\sqrt{M^2-b^2-Q^2}}\bigg\lbrace M+\sqrt{M^2-b^2-Q^2}\bigg\rbrace.
	\end{split}
\end{align}
\\
We shall compute the local mode contribution to logarithmic corrections to the entropy of the non-extremal Kerr-Newman family of  black holes using \cref{local mode}. For the  Kerr-Newmann metric \eqref{metric},
one finds \cite{Henry:2000wd,Cherubini:2002we}:
\begin{align}\label{K}
\begin{split}
  R^{\mu\nu\rho\sigma}R_{\mu\nu\rho\sigma}=&\frac{8}{(r^2+b^2 \cos^2 \psi)^6}\left\{ 6M^2(r^{6}-15b^2r^4 \cos^2 \psi+15 b^4 r^2 \cos^4 \psi  \right.
\\
&\left.-b^6 \cos^6 \psi) -12 M Q^2 r(r^4-10 r^2 b^2 \cos^2 \psi +5b^4 \cos^4 \psi)\right.
\\
& \left. +Q^4(7r^4-34 r^2 b^2 \cos^2 \psi +7 b^4 \cos^4 \psi) \right\},
\\
 R^{\mu\nu}R_{\mu\nu}=&\frac{4Q^4}{(r^2+b^2 \cos^2 \psi)^4}, 
 \\
 \det(\bar{g}_{\mu\nu})=&\bar{g}=(r^2+b^2 \cos^2\psi)^2 \sin^2\psi
 \end{split}
\end{align}
On Accounting $t\to-i \tau$ (Wick rotation of time) where $\tau$ is periodic with $\beta$,
we have integration results for the square of Weyl tensor and Euler density as \cite{Charles:2015nn,Sen:2012dw}
\begin{align}\label{integration 2}
\begin{split}
\int d^4 x \sqrt{\bar{g}} \thickspace W^{\mu\nu\rho\sigma}W_{\mu\nu\rho\sigma}&=
 64 \pi^2+ \frac{\pi \beta Q^4}{b^5 r_{H}^4(b^2+r_{H}^{2})}\Big\{ 3b^5 r_{H}+2b^3r_{H}^3
\\
&\quad+3(b^2-r_{H}^2)(b^2+r_{H}^2)^2 \tan^{-1}\Big(\frac{b}{r_{H}}\Big)+3 b r_{H}^5\Big\},
\\
\int d^4x \sqrt{\bar{g}}E_{4}&=64 \pi^2.
\end{split}
\end{align}
From \cref{integration 2}, one can see that result for the integration of square of Weyl tensor depends upon black hole parameters, i.e., the geometry of background, while the same for the Gauss-Bonnet term is independent of black hole parameter and is a constant. Because of these properties, we have split the term $a_{4}$ in \cref{a4} in two parts, where one is associated with Weyl tensor and the other part is associated with Euler density.  

Now, we will compute the local contribution to entropy of various non-extremal black holes in \say{non-minimal} $\mathcal{N}=1$, $d=4$ EMSGT. It is determined using $a_{4}$ \eqref{a4} along with \eqref{local mode} and \eqref{integration 2}. We get the required logarithmic correction to the entropy in case of Kerr-Newman black hole: 
\begin{align}
	\begin{split}\label{local KN}
 \Delta S_{\text{local}, \text{KN}}
&=\Big\{\frac{17}{384 \pi}\frac{\beta Q^4}{b^5 r_{H}^4(b^2+r_{H}^2)}\Big(3 b^5 r_{H} +2b^3r_{H}^3
\\
&\quad+3(b^2-r_{H}^2)(b^{2}+r_{H}^{2})^2\tan^{-1}\Big(\frac{b}{r_{H}}\Big)+3b r_{H}^5\Big)+\frac{19}{12}\Big\}\ln A_{H} .
\end{split}
\end{align}
For Kerr black hole, setting the limit $Q\to 0$ in \cref{local KN}, one finds the required logarithmic correction to the Bekenstein-Hawking entropy:
\begin{align}\label{local Kerr}
\Delta S_{\text{local}, \text{Kerr}}=\frac{19}{12} \ln A_{H}.
\end{align}
For Reissner Nordstr\"{o}m black hole, $J\to0$ gives $b\to0$. Using $b\to0$ limit in \cref{local KN}, we get the required logarithmic correction to the entropy of Reissner-Nordstr\"{o}m black hole:  
\begin{align}\label{local RN}
\Delta S_{\text{local}, \text{RN}}=\Big(\frac{19}{12}+\frac{17}{60 \pi r_{H}^5}\beta Q^{4}\Big)\ln A_{H}.
\end{align}
Furthermore, setting $Q\to0$ and $b\to0$ in \cref{local KN} gives the result for a Schwarzschild black hole:
\begin{align}\label{local S}
\Delta S_{\text{local}, \text{Schw}}=\frac{19}{12} \ln A_{H}.
\end{align}
We then proceed to determine the zero mode contribution using \cref{zero mode contribution}. Assigning the values of $Y$ as 2 for metric, 1 for gauge field, 1/2 for Dirac field and 3/2 for gravitino field for 4D field theory, the zero mode contribution due to gauge field and Dirac field will vanish because of the factor $(Y-1)$ and $(2Y-1)$\footnote{The fermionic scaling dimensions scale two times because of its spin degeneracy.} for bosonic and fermionic fields in \cref{zero mode contribution} respectively. So, we only have to consider about zero modes associated with metric and gravitino fluctuations in \say{non-minimal} $\mathcal{N}=1$, $d=4$ EMSGT. Again, the black holes in $\mathcal{N}=1$, $d=4$ EMSGT are non-BPS and do not preserve any supersymmetry i.e. $N_{\text{SUSY}}=0$. One can determine the required zero mode contributions to the entropy of Kerr-Newman, Kerr, Reissner-Nordstr\"{o}m and Schwarzschild black holes in \say{non-minimal}  $\mathcal{N}=1$, $d=4$ EMSGT using \cref{zero mode contribution 2}, which is given by 
\begin{align}\label{KNzm}
	\begin{split}
	& \Delta S_{\text{zm,KN}}=\Delta S_{\text{zm,Kerr}}=\frac{1}{2}\left\{(-(3+1)+3 \right\}\ln A_{H}=-\frac{1}{2}\ln A_{H},
	\\
	& \Delta S_{\text{zm,Schw}}=\Delta S_{\text{zm,RN}}=\frac{1}{2}\left\{(-(3+3)+3 \right\}\ln A_{H}=-\frac{3}{2}\ln A_{H}.
	\end{split}
\end{align}
So, the net logarithmic correction to the entropy of non-extremal black holes in \say{non-minimal} $\mathcal{N}=1$ $d=4$ EMSGT will be obtained using \cref{KNzm,local KN,local Kerr,local RN,local S,entropy relation}. The logarithmic corrections to the entropy of non-extremal Kerr-Newman, Kerr, Reissner-Nordstr\"{o}m and Schwarzschild black holes in \say{non-minimal} $\mathcal{N}=1$, $d=4$ EMSGT  are computed as:

\begin{align}\label{entropy KN}
\begin{split}
\Delta S_{\text{BH,KN}}&=
\Big\{\frac{17}{384 \pi}\frac{\beta Q^4}{b^5 r_{H}^4(b^2+r_{H}^2)}\Big(3 b^5 r_{H} +2b^3r_{H}^3
\\
&\quad+3(b^2-r_{H}^2)(b^{2}+r_{H}^{2})^2\tan^{-1}\Big(\frac{b}{r_{H}}\Big)+3b r_{H}^5\Big)
+\frac{13}{12}\Big\}\ln A_{H},
\end{split}
\\
	\begin{split}\label{entropy K}
\Delta S_{\text{BH,Kerr}}&=\frac{13}{12} \ln A_{H},
	\end{split}
\\
	\begin{split}\label{entropy RN}
\Delta S_{\text{BH,RN}}&=\Big(\frac{1}{12}+\frac{17}{60 \pi r_{H}^5}\beta Q^{4}\Big)\ln A_{H},
	\end{split}
\\
	\begin{split}\label{entropy S}
\Delta S_{\text{BH,Schw}}&=\frac{1}{12} \ln A_{H}.
	\end{split}
\end{align}
\section{Concluding remarks}\label{section 6}
To summarize, the whole work is two-fold. First, we have computed the first three Seeley-DeWitt coefficients \eqref{total a2n} for \say{non-minimal} $\mathcal{N}=1$, $d=4$ EMSGT following the field redefinition approach of Seeley-DeWitt expansion \cite{Charles:2015nn}. We reproduced the results for bosonic sector of the theory computed in \cite{Charles:2015nn}. We then evaluated the coefficients for fermionic sector of \say{non-minimal} $\mathcal{N}=1$, $d=4$ EMSGT introducing a particular field redefinition.
We obtained the Seeley-DeWitt coefficients for the theory  by adding both sectors. The results were found to be in perfect agreement with our earlier work \cite{Banerjee:2020wbr}, where the computation is performed via the general approach of Seeley-DeWitt expansion \cite{Vassilevich:2003ll}. 
 It checks the consistency for results of these coefficients. In the second part, we use a particular Seeley-DeWitt coefficient $a_4$ to determine the logarithmic corrections to the Bekenstein-Hawking entropy of non-extremal black holes in \say{non-minimal} $\mathcal{N}=1$, $d=4$ EMSGT. In \cref{GenFrm}, we reviewed the formalism \cite{Sen:2012dw,Charles:2018yey} to compute the logarithmic corrections to the entropy of any arbitrary four dimensional non-extremal black holes. We applied the formalism for Kerr-Newman black hole background and evaluated the local contributions to logarithmic corrections by integrating $a_4$ coefficient over the whole geometry of black hole background. 
  The zero mode contributions to these corrections are determined separately by analyzing the number of zero modes and scaling dimensions of the field fluctuations. An extra factor of $3 \delta$ also appears in zero mode contributions \eqref{zero mode contribution 2} while considering the case for non-extremal black holes. By adding the local and zero mode contributions one finds the total logarithmic corrections to the entropy of non-extremal Kerr-Newman \eqref{entropy KN} black holes.
  The results for Kerr \eqref{entropy K}, Reissner-Nordstr\"{o}m \eqref{entropy RN} and Schwarzschild \eqref{entropy S} black holes are obtained by applying proper limits on the logarithmic correction of Kerr-Newman black hole. The answers for Kerr-Newman and Reissner-Nordstr\"{o}m black holes are found to be dependent on black hole parameters, whereas the same is independent for Kerr and Schwarzschild black holes. The Schwarzschild black holes do not have charge and angular momentum. So one can expect that the logarithmic corrections to the entropy of Schwarzschild black holes will be independent of black holes parameters and thus constant. However, such a priori conclusion cannot be predicted for Kerr black holes, which possess definite angular momentum, and therefore significant and worthy of attention. 
  Again, if we consider the extremal regime within the same theory, it is only Kerr-Newman black hole whose logarithmic correction part depends on the black hole parameter, and others
   are independent of it \cite{Banerjee:2020wbr}.  Our results are new and unique. These are not computed anywhere else in any other known theory. These results put a strong constraint for any other microscopic theory describing the entropy of these black holes. The results may provide directions in the investigation of a conformal field theory describing these black hole's entropies. One can also generalize this $\mathcal{N}=1$, $d=4$ EMSGT by coupling $n_{v}$ number of vector multiplets and $n_{c}$ number of chiral multiplets. As a result, there will be extra contributions in the logarithmic corrections to black hole entropy due to the presence of these extra matter couplings with the black hole background. References \cite{Karan:2020njm, Ferrara:2011qf} compute the logarithmic corrections to the entropy of black holes in a four dimensional matter coupled $\mathcal{N}=1$ EMSGT obtained by truncation of 4D $\mathcal{N}=2$ EMSGT. Even in a macroscopic regime, the other heat kernel methods can be utilized for reproducing these results in \say{non-minimal} $\mathcal{N}=1$, $d=4$ EMSGT.   
One can also observe that the results of non-extremal Kerr-Newman, Kerr and Reissner-Nordstr\"{o}m black holes are positive and greater than that of its extremal counterpart in \cite{Banerjee:2020wbr}. Such a thing is expected in non-extremal limit because we are concerned with the entropy of black hole in whole geometry, unlike the extremal case.
 A proper understanding of these results may provide other valuable insights for studying the Kerr-Newman family of black holes in $\mathcal{N}=1$ EMSGT for future progress. Such work can also be extended by adding a superpotential to this $\mathcal{N}=1$ theory and study these corrections for the Kerr-Newman family of black holes. Again it can also be extended to  Einstein-Maxwell dilaton theory embedded in $\mathcal{N}=1$ theory for studying the logarithmic corrections to the entropy of dyonic black holes following the approach of this paper.     

\acknowledgments
We would like to acknowledge Ashoke Sen, Finn Larsen and Rajesh Kumar Gupta for valuable discussions during the early stages of the research work. We also thank Sudip Karan for reading the manuscript carefully and verifying some parts of the calculations. We are also thankful to the unknown reviewer for valuable remarks, serving better clarity to some parts of the manuscript.

\appendix
\section{Trace calculations}\label{Appendix}
Here, we will present the  trace calculations of $E$, $E^2$ and $\Omega^{\alpha \beta} \Omega_{\alpha \beta}$ for both \textemdash bosonic and fermionic sectors of \say{non-minimal} $\mathcal{N}=1$, $d=4$ EMSGT. These trace results are useful in the computation of $a_0$, $a_2$ and $a_4$. In this computation, we will make use of the following identities whenever required:
\begin{align}
\begin{split}
(D_{\rho}\bar{F}_{\mu\nu})(D^{\rho}\bar{F}^{\mu\nu})&=R_{\mu\nu\theta\phi}\bar{F}^{\mu\nu}\bar{F}^{\theta\phi}-R^{\mu\nu}R_{\mu\nu},
\\
(D_{\mu}\bar{{F}_{\rho}}^{\nu})(D_{\nu}\bar{F}^{\rho\mu})&=\frac{1}{2}(R_{\mu\nu\theta\phi}\bar{F}^{\mu\nu}\bar{F}^{\theta\phi}-R^{\mu\nu}R_{\mu\nu}).
\end{split}
\end{align}
We first begin with bosonic sector of \say{non-minimal} $\mathcal{N}=1$, $d=4$ EMSGT.  
\subsection*{Bosonic sector:}
The expressions of ${E}$ and ${\Omega_{\alpha \beta}}$ for the bosonic sector of $\mathcal{N}=1$, $d=4$ EMSGT are shown in \cref{bosonic E,Bosonic Omega} respectively. The required traces are:
\\~\\
\textbf{Trace of $E$.}
\begin{align}\label{trace E}
\text{tr}\thickspace (E)
= \text{tr}\thickspace({E^{\Phi_{\mu\nu}}}_{\Phi_{\mu\nu}}+{E^{a_\alpha}}_{a_\alpha}+{E^{\Phi_{\mu\nu}}}_{a_{\alpha}}+{E^{a_{\alpha}}}_{\Phi_{\mu\nu}}).
\end{align}
We have
	\begin{equation}\label{componentwise trace E}
	\begin{split}
	\text{tr} \thickspace ({E^{\Phi_{\mu\nu}}}_{\Phi_{\mu\nu}})&=0,
	\\
	\text{tr} \thickspace ({E^{a_{\alpha}}}_{a_{\alpha}})&=6\bar{F}^{\mu\nu}\bar{F}_{\mu\nu},
	\end{split}
	\hspace{0.3in}
	\begin{split}
	\text{tr}\thickspace ({E^{\Phi_{\mu\nu}}}_{a_{\alpha}})&=0,
	\\
	\text{tr} \thickspace ({E^{a_{\alpha}}}_{\Phi_{\mu\nu}})&=0.
	\end{split}
	\end{equation}
	Using \cref{trace E,componentwise trace E}, we get the trace of $E$  \eqref{boson trace}.
\\~\\	
\textbf{Trace of $E^2$.}
\allowdisplaybreaks
\begin{align}\label{trace E2}
\begin{split}
\text{tr}\thickspace (E^2)=& \text{tr}\thickspace( {E^{\Phi_{\mu\nu}}}_{\Phi_{\rho\sigma}}{E^{\Phi_{\rho \sigma}}}_{\Phi_{\mu\nu}} + {E^{a_{\alpha}}}_{a_{\beta}} {E^{a_{\beta}}}_{a_{\alpha}}+{E^{\Phi_{\mu\nu}}}_{\Theta}{E^{\Theta}}_{\Phi_{\mu\nu}}
\\
& +{E^{\Theta}}_{\Phi_{\mu \nu}}{E^{\Phi_{\mu\nu}}} _{\Theta}+{E^{\Phi_{\mu\nu}}}_{a_{\rho}}{E^{a_{\rho}}}_{\Phi_{\mu\nu}}+{E^{a_{\mu}}}_{\Phi_{\rho \sigma}}{E^{\Phi_{\rho \sigma}}}_{a_{\mu}}).
\end{split}
\end{align}
 We have
\begin{align}\label{componentwise E2}
\begin{split}
\text{tr}\thickspace({E^{\Phi_{\mu\nu}}}_{\Phi_{\rho\sigma}}{E^{\Phi_{\rho \sigma}}}_{\Phi_{\mu\nu}} )& = 3 R^{\mu\nu\theta\phi}R_{\mu\nu\theta\phi}-2R^{\mu\nu}R_{\mu\nu},
\\
\text{tr}\thickspace({E^{a_{\alpha}}}_{a_{\beta}} {E^{a_{\beta}}}_{a_{\alpha}})&=9 (\bar{F}^{\mu \nu}\bar{F}_{\mu \nu})^2,
\\
\text{tr}\thickspace({E^{\Phi_{\mu\nu}}}_{\Theta}{E^{\Theta}}_{\Phi_{\mu\nu}})
&=-R^{\mu\nu}R_{\mu\nu},
\\
\text{tr}\thickspace({E^{\Theta}}_{\Phi_{\mu \nu}}{E^{\Phi_{\mu\nu}}} _{\Theta})&=-R^{\mu\nu}R_{\mu\nu},
\\
\text{tr}\thickspace({E^{\Phi_{\mu\nu}}}_{a_{\rho}}{E^{a_{\rho}}}_{\Phi_{\mu\nu}})& =-\frac{3}{2} R^{\mu\nu}R_{\mu\nu}+\frac{3}{2}R_{\mu\nu\theta\phi}\bar{F}^{\mu\nu}\bar{F}^{\theta \phi},
\\
\text{tr}\thickspace({E^{a_{\mu}}}_{\Phi_{\rho \sigma}}{E^{\Phi_{\rho \sigma}}}_{a_{\mu}})&= -\frac{3}{2}R^{\mu\nu}R_{\mu\nu}+\frac{3}{2}R_{\mu\nu\theta\phi}\bar{F}^{\mu\nu}\bar{F}^{\theta\phi}.
\end{split}
\end{align}
From \cref{trace E2,componentwise E2}, we get the $\text{tr}\thickspace (E^2)$ \eqref{boson trace}.
\\~\\
\textbf{Trace of $(\Omega_{\alpha \beta} \Omega^{\alpha \beta})$.}
\begin{align}\label{trace omega2}
\begin{split}
\text{tr}\thickspace(\Omega_{\alpha \beta}\Omega^{\alpha \beta})=&\text{tr}\thickspace\lbrace{(\Omega_{\alpha \beta})^{\Phi_{\mu\nu}\thickspace \Phi_{\rho\sigma}}}{(\Omega^{\alpha \beta})_{\Phi_{\rho \sigma}\thickspace \Phi_{\mu\nu}}}+{(\Omega_{\alpha \beta})^{a_{\mu}}}_{a_\rho}{(\Omega^{\alpha \beta})^{a_{\rho}}}_{a_{\mu}}
\\
&+{(\Omega_{\alpha \beta})^{\Phi_{\mu\nu}}}_ {a_\rho}{(\Omega^{\alpha \beta})^{a_\rho}}_{\Phi_{\mu\nu}}+{(\Omega_{\alpha \beta})^{a_\mu}}_{\Phi_{\rho\sigma}}{(\Omega^{\alpha \beta})^{\Phi_{\rho \sigma}}}_{a_\mu}\rbrace,
\end{split}
\end{align}
where
\begin{align}\label{Term 1}
	\begin{split}
	(\Omega_{\alpha \beta})^{\Phi_{\mu\nu}\thickspace \Phi_{\rho \sigma}}(\Omega^{\alpha \beta})_{\Phi_{\rho \sigma}\Phi_{\mu \nu}} 
	=&\Big\lbrace\underbrace{2{R^{\mu\rho}}_{\alpha \beta}\bar{g}^{\sigma \nu}}_{A_1}+\underbrace{{(\omega_{\alpha})^{\Phi_{\mu\nu}}}_{a_{\theta}}(\omega_{\beta})^{a_{\theta} \thickspace \Phi_{\rho \sigma}}-{(\omega_{\beta})^{\Phi_{\mu\nu}}}_{a_\theta}(\omega_{\alpha})^{a_{\theta}\Phi_{\rho \sigma}}}_{A_2}\Big\rbrace
	\\
	& \times \Big\lbrace\underbrace{\frac{1}{2}\Big({R_{\rho\mu}}^{\alpha \beta}\bar{g}_{\nu\sigma}+{R_{\sigma \mu}}^{\alpha\beta}\bar{g}_{\nu \rho}+{R_{\rho\nu}}^{\alpha \beta}\bar{g}_{\mu\sigma}+{R_{\sigma\nu}}^{\alpha \beta}\bar{g}_{\mu\rho}\Big)}_{B_1}
	\\
	&
	+\underbrace{(\omega^{\alpha})_{\Phi_{\rho \sigma} \thickspace a_{\theta}}{(\omega^{\beta})^{a_{\theta}}}_{\Phi_{\mu\nu}}-(\omega^{\beta})_{\Phi_{\rho\sigma} \thickspace a_{\theta}}{(\omega^{\alpha})^{a_{\theta}}}_{\Phi_{\mu\nu}}}_{B_2}\Big\rbrace,
	\end{split}
\end{align}	
\begin{align}\label{Term 2}
	\begin{split}
	(\Omega_{\alpha \beta})^{a_{\mu} a_{\rho}}(\Omega^{\alpha \beta})_{a_{\rho} a_{\mu}}
	=&\Big\{\underbrace{{R^{\mu \rho}}_{\alpha \beta}}_{A_3}+\underbrace{(\omega_{\alpha})^{a_{\mu}\Phi_{\theta\phi}}{(\omega_{\beta})_{\Phi_{\theta \phi}}}^{a_{\rho}}-(\omega_{\beta})^{a_{\mu}\Phi_{\theta\phi}}{(\omega_{\alpha})_{\Phi_{\theta \phi}}}^{a_{\rho}}}_{A_4}\Big\}
	\\
	& \times \Big\{\underbrace{{R_{\rho \mu}}^{\alpha \beta}}_{B_3}+\underbrace{{(\omega^{\alpha})_{a_{\rho}}}^{\Phi_{\eta \kappa}}(\omega^{\beta})_{\Phi_{\eta \kappa} a_{\mu}}-{(\omega^{\beta})_{a_{\rho}}}^{\Phi_{\eta \kappa}}(\omega^{\alpha})_{\Phi_{\eta \kappa} a_{\mu}}}_{B_4}\Big\},
	\end{split}
\end{align} 
\begin{align}\label{Term 3}
	\begin{split}
	 {(\Omega_{\alpha \beta})^{\Phi_{\mu\nu}}}_{a_{\theta}}{(\Omega^{\alpha \beta})^{a_{\theta}}}_{\Phi_{\mu \nu}}=&\Big\lbrace \underbrace{D_{\alpha}{(\omega_{\beta})^{\Phi_{\mu\nu}}}_{a_{\theta}}-D_{\beta}{(\omega_{\alpha})^{\Phi_{\mu\nu}}}_{a_{\theta}}}_{A_5}\Big\rbrace 
	 \\
	 &\times \Big \lbrace \underbrace{D^{\alpha}{(\omega^{\beta})^{a_{\theta}}}_{\Phi_{\mu\nu}}-D^{\beta}{(\omega^{\alpha})^{a_{\theta}}}_{\Phi_{\mu\nu}}}_{B_5} \Big \rbrace
	 \end{split}
\end{align}
and 
\begin{align}\label{Term 4}
	\begin{split}
{(\Omega_{\alpha \beta})^{a_{\mu}}}_{\Phi_{\rho \sigma}}{(\Omega^{\alpha \beta})^{\Phi_{\rho \sigma}}}_{a_{\mu}}=&\Big\lbrace \underbrace{D_{\alpha}{(\omega_{\beta})^{a_{\mu}}}_{\Phi_{\rho\sigma}}-D_{\beta}{(\omega_{\alpha})^{a_{\mu}}}_{\Phi_{\rho\sigma}}}_{A_6}\Big\rbrace 
\\
&\times \Big \lbrace \underbrace{D^{\alpha}{(\omega^{\beta})^{\Phi_{\rho\sigma}}}_{a_{\mu}}-D^{\beta}{(\omega^{\alpha})^{\Phi_{\rho\sigma}}}_{a_{\mu}}}_{B_6} \Big \rbrace.
\end{split}
\end{align}
We get the following trace results:

	\begin{equation}\label{componentwise trace result omega boson}
	\begin{split}
		\text{tr}\thickspace (A_1 B_1) &=-6R^{\mu\nu\theta\phi}R_{\mu\nu\theta\phi},\\
		\text{tr}\thickspace (A_2 B_2) &=12 R^{\mu\nu}R_{\mu\nu}-24(\bar{F}^{\mu\nu}\bar{F}_{\mu\nu})^2,\\
		\text{tr}\thickspace (A_1 B_2) &=0,\\
		\text{tr}\thickspace (A_2 B_1) &=0,\\
			\text{tr}\thickspace (A_3 B_3)&=-R^{\mu\nu\rho\sigma}R_{\mu\nu\rho\sigma},
	\end{split}
	\hspace{0.3in}
	\begin{split}
		\text{tr} \thickspace (A_{4}B_{4})&=22R^{\mu\nu}R_{\mu\nu}-30(\bar{F}^{\mu\nu}\bar{F}_{\mu\nu})^2,\\
		\text{tr}\thickspace (A_3 B_4)&=2R^{\mu\nu}R_{\mu\nu}
		,\\
		\text{tr}\thickspace (A_4 B_3)&=2R^{\mu\nu}R_{\mu\nu}
		,\\
		\text{tr}\thickspace (A_5 B_5) &=9R^{\mu\nu}R_{\mu\nu}-9R_{\mu\nu\theta\phi}\bar{F}^{\mu\nu}\bar{F}^{\theta \phi},\\
		\text{tr}\thickspace (A_6 B_6) &=9R^{\mu\nu}R_{\mu\nu}-9R_{\mu\nu\theta\phi}\bar{F}^{\mu\nu}\bar{F}^{\theta \phi}.
	\end{split}
\end{equation}
Using \cref{Term 1,Term 2,Term 3,Term 4,componentwise trace result omega boson}, we have
\begin{align}\label{componentwise trace omega}
\begin{split}
\text{tr}\thickspace \lbrace{(\Omega_{\alpha \beta})^{\Phi_{\mu\nu}}}_{ \Phi_{\rho\sigma}}{(\Omega^{\alpha \beta})^{\Phi_{\rho \sigma}}}_{\Phi_{\mu\nu}}\rbrace&=-6R^{\mu\nu\theta\phi}R_{\mu\nu\theta\phi}+12R^{\mu\nu}R_{\mu\nu}-24(\bar{F}^{\mu\nu}\bar{F}_{\mu\nu})^2,
\\
\text{tr}\thickspace \lbrace{(\Omega_{\alpha \beta})^{a_{\mu}}}_{a_\rho}{(\Omega^{\alpha \beta})^{a_{\rho}}}_{a_{\mu}}\rbrace&=-R^{\mu\nu\theta\phi}R_{\mu\nu\theta\phi}+26R^{\mu\nu}R_{\mu\nu}-30(\bar{F}^{\mu\nu}\bar{F}_{\mu\nu})^2,
\\
\text{tr}\thickspace\lbrace{(\Omega_{\alpha \beta})^{\Phi_{\mu\nu}}}_ {a_\rho}{(\Omega^{\alpha \beta})^{a_\rho}}_{\Phi_{\mu\nu}}\rbrace&=9(R^{\mu\nu}R_{\mu\nu}-R_{\mu\nu\theta\phi}\bar{F}^{\mu\nu}\bar{F}^{\theta \phi}),
\\
\text{tr}\thickspace \lbrace{(\Omega_{\alpha \beta})^{a_\mu}}_{\Phi_{\rho\sigma}}{(\Omega^{\alpha \beta})^{\Phi_{\rho \sigma}}}_{a_\mu}\rbrace&=9(R^{\mu\nu}R_{\mu\nu}-R_{\mu\nu\theta\phi}\bar{F}^{\mu\nu}\bar{F}^{\theta \phi}).
\end{split}
\end{align}
Using \cref{trace omega2,componentwise trace omega}, we have the expression for  $\text{tr}(\Omega_{\alpha \beta}\Omega^{\alpha \beta})$ \eqref{boson trace} in the bosonic sector of \say{non-minimal} $\mathcal{N}=1$, $d=4$ EMSGT.

\subsection*{Fermionic sector:}
The expressions of $E$ and ${\Omega_{\alpha \beta}}$ for the fermionic sector of \say{non-minimal} $\mathcal{N}=1$, $d=4$ EMSGT are given in \cref{field redefined E,field redefine Omega}.
\\~\\
\textbf{Trace of $E$.}
\begin{equation}\label{trace E fermion}
\text{tr}\thickspace (E)=\text{tr}\thickspace({E^{\Psi_{\mu}}}_{\Psi_{\mu}}+{E^{\lambda}}_{\lambda}+{E^{\Psi_{\mu}}}_{\lambda}+{E^{\lambda}}_{\Psi_{\mu}}).
\end{equation}
We have
\begin{equation}\label{componentwise trace E fermion}
	\begin{split}
	\text{tr} \thickspace ({E^{\Psi_{\mu}}}_{\Psi_{\mu}})&=-4\bar{F}^{\mu\nu}\bar{F}_{\mu\nu},
	\\
	\text{tr} \thickspace ({E^{\lambda}}_{\lambda})&=-4\bar{F}^{\mu\nu}\bar{F}_{\mu\nu},
	\end{split}
	\hspace{0.3in}
	\begin{split}
	\text{tr}\thickspace ({E^{\Psi_{\mu}}}_{\lambda})&=0,
	\\
	\text{tr} \thickspace ({E^{\lambda}}_{\Psi_{\mu}})&=0.
	\end{split}
	\end{equation}
Using \cref{trace E fermion,componentwise trace E fermion}, we get the $\text{tr}(E)$  expressed in \eqref{trace fermion}.
\\~\\
\textbf{Trace of $E^2$.}

\begin{align}\label{trace E2 fermion}
\begin{split}
\text{tr}\thickspace (E^2)=\text{tr}({E^{\Psi_{\mu}}}_{\Psi_{\nu}}{E^{\Psi_\nu}}_{\Psi_{\mu}}+{E^{\lambda}}_{\lambda} {E^{\lambda}}_{\lambda}+{E^{\Psi_{\mu}}}_{\lambda}{E^{\lambda}}_{\Psi_{\mu}}+{E^{\lambda}}_{\Psi_{\mu}}{E^{\Psi_{\mu}}}_{\lambda}).
\end{split}
\end{align}
We have
\begin{align}\label{componentwise trace E2 fermion}
\begin{split}
\text{tr}\thickspace({E^{\Psi_{\mu}}}_{\Psi_{\nu}}{E^{\Psi_\nu}}_{\Psi_{\mu}})&=5R^{\mu\nu}R_{\mu\nu}+2R^{\mu\nu\theta\phi}R_{\mu\nu\theta\phi}+2(\bar{F}^{\mu\nu}\bar{F}_{\mu\nu})^2,
\\
\text{tr} \thickspace ({E^{\lambda}}_{\lambda} {E^{\lambda}}_{\lambda})&=-4R^{\mu\nu}R_{\mu\nu}+8(\bar{F}^{\mu\nu}\bar{F}_{\mu\nu})^2,
\\
\text{tr}\thickspace({E^{\Psi_{\mu}}}_{\lambda}{E^{\lambda}}_{\Psi_{\mu}})&=R^{\mu\nu}R_{\mu\nu}-R_{\mu\nu\theta\phi}\bar{F}^{\mu\nu}\bar{F}^{\theta\phi},
\\
\text{tr} \thickspace ({E^{\lambda}}_{\Psi_{\mu}}{E^{\Psi_{\mu}}}_{\lambda})&=R^{\mu\nu}R_{\mu\nu}-R_{\mu\nu\theta\phi}\bar{F}^{\mu\nu}\bar{F}^{\theta\phi}.
\end{split}
\end{align}
From \cref{trace E2 fermion,componentwise trace E2 fermion}, we get the $\text{tr}\thickspace(E^2)$ \eqref{trace fermion}.
\\~\\
\textbf{Trace of $(\Omega_{\alpha \beta} \Omega^{\alpha \beta})$.}
\begin{align}\label{trace Omega2 fermion}
\begin{split}
\text{tr} \thickspace (\Omega_{\alpha \beta}\Omega^{\alpha \beta})=&\text{tr}\thickspace \lbrace{(\Omega_{\alpha \beta})_{\Psi_{\mu}}}^{\Psi_{\nu}}{(\Omega^{\alpha \beta})_{\Psi_{\nu}}}^{\Psi_{\mu}}+{(\Omega_{\alpha \beta})_{\lambda}}^{\lambda}{(\Omega^{\alpha \beta})_{\lambda}}^{\lambda}
\\
&+{(\Omega_{\alpha  \beta})_{\Psi_{\mu}}}^{\lambda}{(\Omega^{\alpha \beta})_{\lambda}}^{\Psi_{\mu}}+{(\Omega_{\alpha \beta})_{\lambda}}^{\Psi_{\mu}}{(\Omega^{\alpha \beta})_{\Psi_{\mu}}}^{\lambda}\rbrace
\end{split},
\end{align}
where 
\begin{align}\label{term 1}
	\begin{split}
	&{(\Omega_{\alpha \beta})_{\Psi_{\mu}}}^{\Psi_{\nu}}{(\Omega^{\alpha \beta})_{\Psi_{\nu}}}^{\Psi_{\mu}}
	\\
	=&\bigg\lbrace\underbrace{{{R_{\mu}}^{\nu}}_{\alpha \beta} \mathbb{I}_{4}}_{X_1}+\underbrace{\frac{1}{4}R_{\alpha \beta \xi \kappa}\gamma^{\xi}\gamma^{\kappa}\bar{g^{\nu}}_{\mu}}_{X_{2}}+\Big(\underbrace{{(\omega_{\alpha})_{\Psi_{\mu}}}^{\lambda}{(\omega_{\beta})_{\lambda}}^{\Psi_{\nu}}-{(\omega_{\beta})_{\Psi_{\mu}}}^{\lambda}{(\omega_{\alpha})_{\lambda}}^{\Psi_{\nu}}}_{X_{3}}\Big)\bigg\rbrace
	\\
	&\times \bigg\lbrace\underbrace{{R_{\nu}}^{\mu \alpha \beta}\mathbb{I}_{4}}_{Y_1}+\underbrace{\frac{1}{4}{R^{\alpha\beta}}_{\eta \tau}\gamma^{\eta}\gamma^{\tau}\bar{g^{\mu}}_{\nu}}_{Y_{2}}+\Big(\underbrace{{(\omega^{\alpha})_{\Psi_{\nu}}}^{\lambda}{(\omega^{\beta})_{\lambda}}^{\Psi_{\mu}}-{(\omega^{\beta})_{\Psi_{\nu}}}^{\lambda}{(\omega^{\alpha})_{\lambda}}^{\Psi_{\mu}}}_{Y_{3}}\Big)\bigg\rbrace,
	\end{split}
\end{align}

\begin{align}\label{term 2}
	\begin{split}
	{(\Omega_{\alpha \beta})_{\lambda}}^{\lambda}{(\Omega^{\alpha \beta})_{\lambda}}^{\lambda}=&\bigg\lbrace\underbrace{\frac{1}{4}\gamma^{\theta}\gamma^{\phi}R_{\alpha \beta \theta \phi}}_{X_{4}}+\underbrace{{(\omega_{\alpha})_{\lambda}}^{\Psi_{\mu}}{(\omega_{\beta})_{\Psi_{\mu}}}^{\lambda}-{(\omega_{\beta})_{\lambda}}^{\Psi_{\mu}}{(\omega_{\alpha})_{\Psi_{\mu}}}^{\lambda}}_{X_{5}}\bigg\rbrace
	\\
	& \times\bigg\lbrace\underbrace{\frac{1}{4}\gamma^{\xi}\gamma^{\kappa}{R^{\alpha \beta}}_{\xi \kappa}}_{Y_{4}}+\underbrace{{(\omega^{\alpha})_{\lambda}}^{\Psi_{\nu}}{(\omega^{\beta})_{\Psi_{\nu}}}^{\lambda}-{(\omega^{\beta})_{\lambda}}^{\Psi_{\nu}}{(\omega^{\alpha})_{\Psi_{\nu}}}^{\lambda}}_{Y_{5}}\bigg\rbrace,
	\end{split}
\end{align}
\begin{align}\label{term 3}
	{(\Omega_{\alpha  \beta})_{\Psi_{\mu}}}^{\lambda}{(\Omega^{\alpha \beta})_{\lambda}}^{\Psi_{\mu}}=\Big (\underbrace{D_{\alpha}{(\omega_{\beta})_{\Psi_{\mu}}}^{\lambda}-D_{\beta}{(\omega_{\alpha})_{\Psi_{\mu}}}^{\lambda}}_{X_{6}} \Big) \times \Big(\underbrace{D^{\alpha}{(\omega^{\beta})_{\lambda}}^{\Psi_{\mu}}-D^{\beta}{(\omega^{\alpha})_{\lambda}}^{\Psi_{\mu}}}_{Y_{6}}\Big)
\end{align}
and 
\begin{align}\label{term 4}
	{(\Omega_{\alpha \beta})_{\lambda}}^{\Psi_{\mu}}{(\Omega^{\alpha \beta})_{\Psi_{\mu}}}^{\lambda}=\Big(\underbrace{D_{\alpha}{(\omega_{\beta})_{\lambda}}^{\Psi_{\mu}}-D_{\beta}{(\omega_{\alpha})_{\lambda}}^{\Psi_{\mu}}}_{X_{7}}\Big)\times \Big (\underbrace{D^{\alpha}{(\omega^{\beta})_{\Psi_{\mu}}}^{\lambda}-D^{\beta}{(\omega^{\alpha})_{\Psi_{\mu}}}^{\lambda}}_{Y_{7}} \Big).
\end{align}
We get the following trace results:
	\begin{equation}\label{componentwise trace result omega fermion}
	\begin{split}
		\text{tr}\thickspace (X_1 Y_1) &=-4R^{\mu\nu\theta\phi}R_{\mu\nu\theta\phi},\\
		\text{tr}\thickspace (X_2 Y_2) &=-2R^{\mu\nu\theta\phi}R_{\mu\nu\theta\phi},\\
		\text{tr}\thickspace (X_1 Y_2) &=0,\\
		\text{tr}\thickspace (X_2 Y_1) &=0,\\
		\text{tr}\thickspace (X_1 Y_3) &=4R^{\mu\nu}R_{\mu\nu}-4R_{\mu\nu\theta\phi}\bar{F}^{\mu\nu}\bar{F}^{\theta \phi},\\
		\text{tr}\thickspace (X_3 Y_1) &=4R^{\mu\nu}R_{\mu\nu}-4R_{\mu\nu\theta\phi} \bar{F}^{\mu\nu} \bar{F}^{\theta \phi},\\
		\text{tr}\thickspace (X_2 Y_3)&=-2R^{\mu\nu}R_{\mu\nu},\\
		\text{tr}\thickspace (X_3 Y_2)&=-2R^{\mu\nu}R_{\mu\nu},\\
	\end{split}
	\hspace{0.3in}
	\begin{split}
		\text{tr}\thickspace (X_3 Y_3)&=6R^{\mu\nu}R_{\mu\nu}-36 (\bar{F}^{\mu\nu}\bar{F}_{\mu\nu})^2,\\
		\text{tr} \thickspace (X_{4}Y_{4})&=-\frac{1}{2}R^{\mu\nu\theta\phi}R_{\mu\nu\theta\phi},\\
		\text{tr} \thickspace (X_4 Y_{5})&=-2R^{\mu\nu}R_{\mu\nu},\\
		\text{tr} \thickspace (X_5 Y_{4}) &=-2R^{\mu\nu}R_{\mu\nu},\\
		\text{tr}\thickspace (X_5 Y_5) &=8R^{\mu\nu}R_{\mu\nu}-24(\bar{F}^{\mu\nu}\bar{F}_{\mu\nu})^2,\\
		\text{tr}\thickspace (X_6 Y_6) &=10 R_{\mu\nu\theta\phi}\bar{F}^{\mu\nu}\bar{F}^{\theta \phi}-10 R^{\mu\nu}R_{\mu\nu},\\
		\text{tr}\thickspace (X_7 Y_7) &=10 R_{\mu\nu\theta\phi}\bar{F}^{\mu\nu}\bar{F}^{\theta \phi}-10 R^{\mu\nu}R_{\mu\nu}.
	\end{split}
\end{equation}
From \cref{term 1,term 2,term 3,term 4,componentwise trace result omega fermion}, we have
\begin{align}\label{componentwise trace Omega2 fermion}
	\begin{split}
		\text{tr}\thickspace \lbrace {(\Omega_{\alpha \beta})_{\Psi_{\mu}}}^{\Psi_{\nu}}{(\Omega^{\alpha \beta})_{\Psi_{\nu}}}^{\Psi_{\mu}}\rbrace &= -6R^{\mu\nu\theta\phi}R_{\mu\nu\theta\phi}+10R^{\mu\nu}R_{\mu\nu}
		\\
		&\quad-8R_{\mu\nu\theta\phi}\bar{F}^{\mu\nu}\bar{F}^{\theta\phi}-36(\bar{F}^{\mu\nu}\bar{F}_{\mu\nu})^2, 
		\\
		\text{tr}\thickspace \lbrace {(\Omega_{\alpha \beta})_{\lambda}}^{\lambda}{(\Omega^{\alpha \beta})_{\lambda}}^{\lambda}\rbrace&=-\frac{1}{2}R^{\mu\nu\theta\phi}R_{\mu\nu\theta\phi}+4R^{\mu\nu}R_{\mu\nu}-24(\bar{F}^{\mu\nu}\bar{F}_{\mu\nu})^2, 
		\\
		\text{tr}\thickspace \lbrace {(\Omega_{\alpha  \beta})_{\Psi_{\mu}}}^{\lambda}{(\Omega^{\alpha \beta})_{\lambda}}^{\Psi_{\mu}} \rbrace&=10(R_{\mu\nu\theta\phi}\bar{F}^{\mu\nu}\bar{F}^{\theta\phi}-R^{\mu\nu}R_{\mu\nu}),
		\\
		\text{tr}\thickspace \lbrace {(\Omega_{\alpha \beta})_{\lambda}}^{\Psi_{\mu}}{(\Omega^{\alpha \beta})_{\Psi_{\mu}}}^{\lambda}\rbrace&=10(R_{\mu\nu\theta\phi}\bar{F}^{\mu\nu}\bar{F}^{\theta\phi}-R^{\mu\nu}R_{\mu\nu}).
	\end{split}
\end{align}
Using \cref{componentwise trace Omega2 fermion,trace Omega2 fermion}, one can get the $\text{tr}\thickspace (\Omega_{\alpha \beta}\Omega^{\alpha \beta})$ \eqref{trace fermion} for the fermionic sector of \say{non-minimal} $\mathcal{N}=1$, $d=4$ EMSGT.

\pagebreak


\end{document}